\voffset=.4in
\magnification=1200
\baselineskip=20pt
\parindent=.5cm
\font\smallrm=cmr8
\font\smallit=cmti8
\font\smallsl=cmsl8
\font\smallbf=cmbx7 scaled \magstep1

\def\IC{\relax\leavevmode\hbox{\,$\inbar\kern-.3em{\rm C}$}}
\def\IR{\relax{\rm I\kern-.18em R}}
\def\linadjp#1
{\normalbaselines  \multiply\baselineskip#1 \divide\baselineskip100}

\def\smfonts{\let\it\smallit \let\sl\smallsl \let\bf\smallbf
\let\rm\smallrm}

\newfam\bmatfam
\font\tenbi=cmmib10    \font\tenbsy=cmbsy10
\font\sevenbi=cmmib7   \font\sevenbsy=cmbsy7
\font\fivebi=cmmib5    \font\fivebsy=cmbsy5
\font\tenbit=cmbxti10  \font\tenbsl=cmbxsl10
\def\boldmath{ \textfont0=\tenbf  \scriptfont0=\sevenbf
\scriptscriptfont0=\fivebf
\textfont1=\tenbi
\scriptfont1=\sevenbi
\scriptscriptfont1=\fivebi
\textfont2=\tenbsy
\scriptfont2=\sevenbsy
\scriptscriptfont2=\fivebsy
\textfont\itfam=\tenbit
\textfont\slfam=\tenbsl}
\def\bfall{\boldmath \let\tenrm\tenbf \let\tenit\tenbit
\let\tensl\tenbsl \rm}

\parindent=1cm
\font\smallrm=cmr8
\font\smallit=cmti8
\font\smallsl=cmsl8
\font\smallbf=cmbx7 scaled \magstep1

\def\linadjp#1
{\normalbaselines  \multiply\baselineskip#1 \divide\baselineskip100}

\def\smfonts{\let\it\smallit \let\sl\smallsl \let\bf\smallbf
\let\rm\smallrm}

\newfam\bmatfam
\font\tenbi=cmmib10    \font\tenbsy=cmbsy10
\font\sevenbi=cmmib7   \font\sevenbsy=cmbsy7
\font\fivebi=cmmib5    \font\fivebsy=cmbsy5
\font\tenbit=cmbxti10  \font\tenbsl=cmbxsl10
\def\boldmath{ \textfont0=\tenbf  \scriptfont0=\sevenbf
\scriptscriptfont0=\fivebf
\textfont1=\tenbi
\scriptfont1=\sevenbi
\scriptscriptfont1=\fivebi
\textfont2=\tenbsy
\scriptfont2=\sevenbsy
\scriptscriptfont2=\fivebsy
\textfont\itfam=\tenbit
\textfont\slfam=\tenbsl}
\def\bfall{\boldmath \let\tenrm\tenbf \let\tenit\tenbit
\let\tensl\tenbsl \rm}
\def\1{\vskip 1cm}
\def\hb{\hfill\break}
\def\n{\noindent}

\font\title=cmbx10 scaled \magstep2
\font\dept=cmr10 scaled \magstep1
\hfuzz=30pt

{\nopagenumbers
\item{}
\vskip 2 true cm
\centerline{\title{Time Asymmetric Quantum Physics}}
\vskip .5 true cm
\centerline{\title {A.~Bohm}}
%\vskip .2 true cm
\centerline{\dept{Physics Department}}
\centerline{\dept{The University of Texas at Austin}}
\centerline{\dept{Austin, Texas 78712}}
\vfill
{\smfonts\rm e-mail: bohm@physics.utexas.edu}\eject}

\pageno=1
\item{}
\vskip 2 true cm
\centerline{\noindent{\bf Abstract}}
\vskip 0.5 true cm
\centerline{\vbox{\hsize=5 true in  
\noindent Mathematical and phenomenological arguments in favor of
asymmetric time evolution of micro-physical
states are presented.}}

\vfil\eject

\noindent{\bf 1.  Introduction}

\n Standard quantum mechanics in Hilbert space ${\cal H}$ is a
time symmetric theory with a time symmetric dynamical
(differential) equation and time symmetric boundary
conditions.  This is in contrast to many time asymmetric
phenomena observed in classical and also in quantum physics.  Of
the latter we want to discuss in this article two examples, the
decay of a quasi-stable particle [1] and the expanding universe as
a whole when considered as a closed (without extrinsic
influences [2]) quantum systems [3].  

In classical physics solutions of time-symmetric dynamical 
equations with time asymmetric boundary 
conditions come in pairs, e.g., big bang --
big crunch in general relativity or retarded -- advanced in
electromagnetism.  With the choice of the boundary condition,
one out of the two time-asymmetric solutions is selected.  
The Hilbert space
theory of quantum mechanics does not allow such time-asymmetric
formulations.  In the Hilbert space  formulation of quantum mechanics
the 
space-time transformations (e.g., Galilean transformations,
Poincar\'e transformations) are described by a unitary group
representation in the Hilbert space  ${\cal H}$.  Thus the time evolution
is unitary and reversible, and it is given by
$U^\dagger(t)=\exp(-iHt),-\infty<t<\infty$.  This is the
consequence of a series of mathematical theorems which follow
from the mathematical properties - specifically the
topological completeness - of the Hilbert space; they are
listed in the Appendix A.  These theorems in particular exclude
the existence of non-zero probabilities which are zero before a
given finite time $t_0$ ($t_0\not=-\infty$), which is the time
at which the quasi-stable particle had been produced or the time
of the big bang in the two examples of this article. The  decay of
resonances and the quantum theory of our universe can therefore
not be described consistently in the mathematical theory using
the Hilbert space.

Disregarding Hilbert space mathematics, in scattering theory
one arrived in a heuristic way at a pair of time asymmetric
boundary conditions by choosing in- and out-plane wave
``states" $|E^+\rangle$
and $|E^-\rangle$ which have their origin in the $\epsilon=+0$
and $\epsilon=-0$ of the Lippmann-Schwinger 
equation [4], cf.~Appendix B.  Still, the widespread opinion 
remained that asymmetric or irreversible time evolution 
of closed quantum mechanical systems is impossible.

It could have been that historically the analogy to classical
mechanics was the origin of this belief, though the time
evolution for the Schr\"odinger equation could have as well been
discussed in analogy to the electromagnetic waves, and for those
the radiation arrow of time was well accepted (and by some even
considered as fundamental [5]). However, the reversibility of the
Hamiltonian generated time evolution in von Neumann's [6]
Hilbert space quantum theory must have been a decisive factor
for the longevity of this belief.

Already the Dirac [7] kets $|E\rangle$,
$0\leq E\ <\ \infty$, are not elements of the Hilbert
space but generalized eigenvectors and required the extension
of the Hilbert space ${\cal H}$ to the Rigged Hilbert Space
$\Phi\subset{\cal H}\subset\Phi^\times$ [8], where
$\Phi$ is a linear scalar product space of well-behaved vectors
$\phi\in\Phi$ (represented by smooth $etc.$ functions $\langle
E|\phi\rangle$) and $\Phi^\times\ni|E\rangle$ is the space of
its antilinear functionals.  In the $S$-matrix element
[cf.~Appendix B]
$$
(\psi^{\rm out},S\phi^{\rm in})=(\psi^{-},\phi^{+})
=\sum_{bb'}\int_{0}^{\infty}dE
\langle \psi^{-}|b,E^{-}\rangle\langle b|S(E)|b'\rangle
\langle^{+}b',E|\phi^{+}\rangle\leqno(1.1)$$
appear the Dirac ``scattering states" $|E^\pm\rangle$ which
are obtained from $|E\rangle$ by the Lippmann-Schwinger
equation.  In order to analytically continue to the resonance pole 
$z_R=E_R-i\Gamma/2$ of the $S$-matrix $\langle b|S(z)|b^{\prime}\rangle$
the set of in-states $\{\phi^+\}\equiv\Phi_-$ and out-states
$\{\psi^-\}\equiv\Phi_+$ must additionally have some
analyticity property.  In order to get a Breit-Wigner energy
distribution for the pole term we postulate that the energy
wave functions
$\langle^-E|\psi^-\rangle$ and
$\langle^+E|\phi^+\rangle$ are well-behaved Hardy class
functions of the upper and lower half-plane in the
second sheet of the energy surface of the $S$-matrix.

The analytically continued Dirac kets $|E^-\rangle\in\Phi^\times_+$ 
of the Lippmann-Schwinger
equations become -- using the Cauchy formula -- at the
resonance pole
$z_R=E_R-i\Gamma/2$ the Gamow kets
$|z_R^-\rangle\in\Phi^\times_+$.  The time asymmetric semigroup
evolution of these Gamow kets 
$$e^{-iH^\times t}\left|_{\atop t\geq 0}|z_R^-\rangle\right.=
e^{-iE_Rt}e^{-\Gamma/2t}|z_R^-\rangle,\quad\rm{for}\quad t\geq 0\ {\rm
only},  
\leqno(1.2)$$ 
is then derived as 
a mathematical consequence of the structure of the Rigged
Hilbert Space
$\Phi_+^{\times}\supset{\cal H}\supset\Phi_+$ of Hardy class [9] (in
the same way as the time symmetric unitary group evolution given by
$e^{-iH^{\dagger}t}$ $-\infty<t<+\infty$, is a mathematical 
consequence of the Hilbert space structure).

Thus asymmetric time evolution would be a natural property of
quantum mechanical states represented by the vector
$|z_R^-\rangle$ and other elements of the space
$\Phi^\times_+$.  In this article we want to discuss the
phenomenological evidence for such states and the experimental
conditions and phenomenological reason for the
asymmetric time evolution.

In section 2 we review the basic concepts of quantum physics in
a way that shows which mathematical properties are important
for quantum mechanical calculations and which are idealizations
and not directly obtainable from experimental data.  We also
argue that experimental observations involve a time asymmetry,
the preparation $\Rightarrow$ registration arrow of time. 
We then give two
examples of quantum mechanical states with asymmetric time
evolution, the quasi-stable particle and the universe
considered as a closed quantum system, and discuss their common
features.  In section 3 we provide the mathematical theory for
time asymmetric quantum mechanics and give some of its result.
In section 4 we discuss an example of a state with an arrow of
time prepared in a laboratory experiment; we compare the
concept of its preparation time with the initial time for the
state of the quantum universe.

\vfil\eject
\noindent{\bf 2)  Calculational Methods, Mathematical
Idealizations and Experimental Observations}

In quantum theory one has states and observables. States are described by
density or statistical operators and conventionally
denoted by $\rho$ or $W$; for pure states
vectors $\phi$ are used.  Observables are
described by operators $A(=A^\dagger),\Lambda,\ P(=P^2)$, but we will
also use vectors $\psi$ to describe a state
$P$ if $P=|\psi\rangle\langle\psi|$.

The vectors $\phi,\psi$ are elements of a vector space $\Phi$ with a scalar
product,
denoted $(\cdot,\cdot)$ or $\langle\cdot|\cdot\rangle$.
The operators $A,\ \Lambda$, are elements of the algebra of linear
operators ${\cal A}$
in $\Phi$.  The linear space $\Phi$, though
often called a Hilbert space, is mostly treated like a pre-Hilbert space,
i.e., without a topology (or without a
definition of convergence) and it is not topologically complete.
If we want to emphasize that $\Phi$ has no topology we denote it by
$\Phi_{\rm alg}$.

Each ``kind" of quantum physical system is associated to a
space $\Phi$.

In experiments,
the state $W$ (or the pure (idealized) state $\phi$) is prepared by a
preparation apparatus 
and the observable $A$
(or the idealized observable $\psi$) is registered by a registration
apparatus (e.g., a
detector).
The fundamental aspect of the new theory presented here is to clearly
distinguish between states (e.g.,
in-states $\phi^+$ of a scattering experiment) and observables (e.g., 
detected out-states $\psi^-$ of a
scattering experiment), cf.~Appendix B.
\vskip 25pt 
The measured (or registered)
quantities are ratios of (usually) large numbers,
the detector counts.  They are
interpreted as probabilities, e.g.,  as the
probability to measure the observable $\Lambda$ in the state $W$ at
the time
$t$,  which is denoted by ${\cal{P}}_W(\Lambda(t))$.

The probabilities are calculated in theory as the
scalar product or, in the general case, as
the trace. This is shown in relations (2.1a) and (2.1b), 
below where $\approx$
indicates the
equality between the experimental and the
theoretical quantities and $\equiv$ is the mathematical 
definition of the
theoretical probabilities in terms of
the quantities of the space $\Phi$ (which is not yet 
completely  defined):

$$
N_i/N\approx{\cal P}_\phi(P)\equiv|\langle\psi|\phi\rangle|^2\leqno(2.1a)
$$
$$
N(t)/N\approx{\cal P}_W(\Lambda(t))\equiv
Tr(\Lambda(t)W_0)=Tr(\Lambda_0W(t))\leqno(2.1b)
$$

The parameter $t$ in (2.1b) is the continuous time parameter and the
observable $\Lambda$, or the state $W$, are
``continuous" functions of time (with
$W_0=W(t=0$)).  Thus, ${\cal P}_W(\Lambda(t)$) is
thought of as a continuous function of $t$.  But $N(t)$ is the number of
counts in the time interval between $t=0$
and $t$, which is an integer.  Thus the right hand side of $\approx$
in (2.1b) changes continuously in $t$,
but the left hand side can only change in steps of rational numbers.  This
shows that the continuity of
${\cal P}_W(\Lambda(t))$ or of
$|\langle\psi(t)|\phi\rangle|^2=|\langle\psi|\phi(t)\rangle|^2$ as a
function of $t$, and similar topological questions, are not
directly experimentally testable.

For more general observables $A$, which are
expressed in terms of the orthogonal
projection operators $P_i\ (P_iP_j=\delta_{ij}P_j$) as
$$
A=\sum^\infty_{i=1}a_iP_i,\leqno(2.2)
$$
where $a_i$ are the eigenvalues of $A$, 
the probabilities are measured as the average value $\sum_{i=1}^{\rm finite}
a_i{N_i\over N}$. Here the sum is
finite since an experiment can give only a finite number of data.  In the
comparison between theory and
experiment this finite sum is represented by the infinite sum obtained from
(2.2), thus
$$
\sum^{\rm finite}_{i=1}a_i{N_i\over N}\approx{\cal P}(A)=
\sum_{i=1}^{\infty}a_i{\cal P}(P_i).\leqno(2.3)
$$
This also shows that the meaning of such topological notions as the
convergence of infinite sequences (of e.g.,
partial sums of the right hand side of (2.2) and (2.3)) cannot be established
directly from the experimental data on the left hand side
of (2.3), which provides
only a finite sequence.  Thus
the definition of convergence of
infinite sequences in $\Phi$, i.e. the topology of the space $\Phi$,is a
mathematical idealization.
If one wants a complete mathematical theory one needs to make this
mathematical idealization and choose a
topology for the space $\Phi$. Usually, for many
practical calculations in physics, one does not worry about the
completeness and uses
instead some calculational rules.

To obtain the rules for calculating the trace and the scalar product on the
right hand side of (2.1) one starts with a
basis vector decomposition for the state vector $\phi\in\Phi$ using a
discrete set of eigenvectors
$|i\rangle=|\lambda_i\rangle$ of an observable (often the Hamiltonian)
with eigenvalues $\lambda_i$.
$$
\phi=\sum|i\rangle\langle i|\phi\rangle\leqno(2.4)
$$
Often, following Dirac [7], one uses a continuous set of eigenvectors
$|\lambda\rangle$ (Dirac kets) and
writes:
$$
\phi=\int d\lambda|\lambda\rangle\langle\lambda|\phi\rangle\leqno(2.5)
$$
The trace, scalar product, etc., are then calculated as
$$
\leqalignno
{{\rm Tr}(\Lambda W)&=\sum_i^\infty\langle i|\Lambda
W|i\rangle\quad\quad{\rm or,}\ (2.6b)\quad
{\rm Tr}(\Lambda W)=\int d\lambda\langle\lambda|\Lambda
W|\lambda\rangle&(2.6a)\cr
|\langle\psi|\phi\rangle|^2&=
\left|\sum_{i=1}^\infty\langle\psi|i\rangle\langle
i|\phi\rangle\right|^2\quad {\rm or,}\ (2.7b)\quad 
|\langle\psi|\phi\rangle|^2 =
\left|\int d\lambda\langle\psi|\lambda\rangle\langle\lambda|\phi\rangle
\right|^2&(2.7a)\cr}
$$
In practical calculations the convergence of infinite sums and the meaning
of integration (Lebesgue versus
Riemann) are usually not considered.  Often one truncates to 
finite (e.g.,
two) dimensions such that of the sums in (2.6a) and (2.7a) one 
retains only a finite number of terms.  If one has a
complete mathematical theory one can define the meaning of the infinite
sums in (2.6a) (2.7a) and the meaning
of the integrals in (2.6b), (2.7b) and prove (2.4) and (2.5).  For instance
one can choose for $\Phi$ the
Hilbert space ${\cal H}$, in which case (2.4) but not (2.5) can be proven.
Or one can choose for $\Phi$ a
complete space with some locally convex, nuclear topology and its space of
continuous functionals $\Phi^\times$ to obtain
a Gelfand triplet $\Phi\subset{\cal H}\subset\Phi^\times$. Then
the kets are
$|\lambda\rangle\in\Phi^\times$
and one can
prove the Dirac basis vector expansion
(2.5) as the Nuclear Spectral Theorem.
\vskip 28pt 
Time evolution, i.e., the dynamics of a quantum physical system, is
given by the  Hamilton 
operator $H$ of the system.  ($H$ is always assumed to
be (essentially) self-adjoint, $\bar H=H^\dagger$, and semibounded). 
The dynamical equation is 
the von
Neumann or Schroedinger equation:
$$
\leqalignno{
{{\partial W(t)}\over{\partial t}}&={{i}\over{\hbar}}\left[H,W(t)\right];\qquad
i\hbar{{\partial\phi(t)}\over{\partial t}}=H^\dagger\phi(t)&(2.8)\cr
\noalign{\vskip7pt}
\phi(t&=0)=\phi_0\cr}
$$
Equivalently, one gives the time evolution in the Heisenberg picture by
$$
\eqalign{
{{\partial\Lambda(t)}\over{\partial t}}
=-{{i}\over{\hbar}}\left[H,\Lambda(t)\right];\qquad
 i\hbar{{\partial\psi(t)}\over{\partial t}}&=-H\psi(t)\cr
&\psi(t=0)=\psi_0\cr}
$$
In a time symmetric theory, that means if one uses for the time symmetric
differential equation (2.8) {\it also} time
symmetric boundary conditions, then, one obtains the following solutions of
(2.8):
$$
W(t)=e^{-iHt}W_0e^{iHt},\ {\rm where}\ -\infty<t<\infty,\leqno(2.9)
$$
$$
\phi(t)=U^{\dagger}(t)\phi_0=e^{-iHt}\phi_0,\qquad-\infty<t<\infty\leqno(2.10)
$$
or, in the Heisenberg picture,
$$
\Lambda(t)=e^{iHt}\Lambda_0e^{-iHt},\qquad {\rm where}\
-\infty<t<\infty,\leqno(2.11)   
$$
Here $\Lambda_0\equiv\Lambda(t=0),\ W_0\equiv W(t=0)$.

On the other hand if one just starts with the differential equations (2.8)
and postulates the Hilbert
space topology, $\phi(t)\in{\cal H}$, then the above unitary group
evolution is the
only possible solution of the dynamical equations (this follows from some
theorems of
Gleason and Stone (Appendix A)).  This
means time asymmetric boundary conditions which could result in an
irreversible time evolution are not mathematically allowed in a quantum
theory in the Hilbert space
${\cal H}$. The assumption $\phi(t)\in{\cal H}$ always leads to the time
evolution (2.10) given by the
unitary group $U(t)$ which has always an inverse $U(-t)$.

\n Inserting (2.9), (2.10) or (2.11) into the right hand side of (2.1), the 
probability ${\cal P}(t)={\rm Tr}(\Lambda W(t))$ can be 
calculated at any time $t_0+t$ or $t_0-t$.

In contrast to the results calculated with (2.9), the probabilities
${\cal P}(t)$ cannot be observed at
any arbitrary positive or negative time $t$.  The reason is
the following:\hfill\break
{\it A state needs to be prepared before an
observable can be measured, or registered  in it}.\hb  
We call this truism
the preparation $\Longrightarrow$ registration
arrow of time [18]; it is an expression of causality.  Let $t_0(=0)$ be the
time at which the state has been
prepared.  Then, ${\cal P}(\Lambda(t))$ is measured as the ratio of
detector counts
$$\leqalignno{
{\cal P}^{\rm exp}_W(\Lambda(t))&\approx {{N(t)}\over{N}},&(2.12a)\cr
&{\rm for}\ t>t_0=0,&(2.12b)\cr}
$$
If there are some detector counts before $t=t_0$, they are discounted as
noise because the experimental
probabilities
$$
{\rm can}\  {\it not}\  {\rm fulfill}\quad {\cal P}^{\rm
exp}_W(\Lambda(t))\not\approx 0,  \qquad {\rm for}\
t<t_0=0,\leqno(2.13)
$$
Though in the Hilbert space theory ${\cal P}_W(\Lambda(t))={\cal
P}_{W(t)}(\Lambda)$ can be calculated at positive or
negative values of $t-t_0$ using unitary group evolution (2.9), an
experimental meaning can be given to ${\cal
P}_W(\Lambda(t))$ only for $t>t_0$.  

In some cases (e.g., stationary
states, cyclic evolutions), it should not matter at what time ${\cal
P}_{W(t)}(\Lambda)$  is calculated because one can extrapolate to negative
values of $t$.

The physical question is:  Are there quantum physical
states in nature that evolve only into the positive direction of time,
$t>t_0$, and for which one therefore cannot extrapolate to negative
values of $t-t_0$?  If
there are such states, pure states or
mixtures, they cannot be described by the standard Hilbert space 
quantum theory,
because of the unitary group time
evolution (2.9) and (2.10), which is a mathematical consequence of the specific
(topological, not algebraic)
structure of the Hilbert space.

Two prominent examples of states with an asymmetric time evolution,
$t>t_0$, are the decaying states (in all
areas of physics, relativistic or non-relativistic) and our universe as a
whole, considered as a quantum physical
system.

1.)  Decaying states and resonances are often thought of as something
complicated, because in the Hilbert space there does not
exist a vector that can describe them in the same way as stable
states are described by energy eigenvectors.  However,
empirically, quasi-stable particles are not qualitatively different
from stable particles; they differ only quantitatively by a 
non-zero value of the
width $\Gamma$. Stability or the value of lifetime is not taken as a criterion
of elementarity, at least not by the practitioners [10].
A particle decays if it can and it remains stable if
selection rules for some quantum numbers prevent it from
decaying. Therefore, stable and quasi-stable states should be described on the
same footing, e.g., define both by a pole of the S-matrix at the
position $z_R=E_R-i\Gamma/2$, or/and as a generalized eigenvector with
eigenvalue $z_R$ (with $\Gamma=0$ for stable particles).
Since the latter is not possible in the Hilbert space, one
devises ``Effective Theories'' in order to obtain a state vector 
description of quasi-stable states.

Phenomenological effective theories have been enormously
successful. They describe
resonances in a finite dimensional
space as eigenvectors of the
``effective Hamiltonian'' with complex eigenvalue $(E_R-i\Gamma/2)$,
where $E_R$=resonance energy, $\hbar/\Gamma$=life time, and their time
evolution is given by the exponential law. The common feature of
these approximate methods is the omission of a continuous sum; the
infinite dimensional theory is truncated to a finite (e.g., two)
dimensional effective theory.
Examples of this approach are:  The approximate method of Weisskopf and
Wigner and of
Heitler for atomic decaying states [11];
the Lee-Oehme-Yang effective two dimensional theory
for the neutral Kaon system [12]; and many more finite dimensional models
with non-Hermitian diagonalizable Hamiltonian matrices in nuclear
physics [14]. Also non-diagonalizable finite dimensional
Hamiltonians were discussed [13].
In the Hilbert space framework ``there does not exist .~.~., a rigorous
theory to which these methods can be considered as approximations''
[15]. 

The decay of a quantum physical system, e.g., the transition of an 
excited state of a molecule into its ground
state or the decay of an elementary particle [16] 
is a profoundly irreversible
process.  Therefore we should like to
introduce state vector $|F\rangle,\ |\psi^G\rangle=|E_R~-~i\Gamma/2\rangle$
or state operators
$W^G(t)=|F\rangle\langle F|$, for which the time evolution is asymmetric
and for which the theoretical
probabilities Tr($\Lambda W^G(t))$ can be calculated 
for $t>t_0=0$ only.

This means we have
to generalize the unitary group evolution (2.9), (2.10) with $-\infty<
t<\infty$ to a semigroup evolution with $0\leq t<\infty$.  This is
accomplished by seeking
solutions of the time symmetric dynamical equations (2.8)
with time {\it asymmetric} boundary conditions.
Since in Hilbert space quantum mechanics semigroup evolution is not possible,
we seek a semigroup solution $F(t)$ to the quantum mechanical Cauchy problem
(2.8) with Hamiltonian $H^\times_+$
where $F(t)$ is an element of a larger space in which $\cal H$ is
dense and  which we denote by
$\Phi^\times_+\supset{\cal H}$, i.e.,
the Hamiltonian $H^\times_+$ is the uniquely defined extension of the 
Hilbert space Hamiltonian $H^\dagger$ to this space $\Phi^\times_+$. 
Thus the dynamical equation (2.8) is:
$$
i\hbar{{\partial F(t)}\over{\partial t}}=H_+^\times F(t)\leqno(2.14)
$$
with the initial data  $F(t=0)=F^-_0\in\Phi^\times_+$,  
and the solution is given by the 
semigroup\footnote{$^1$}
{\rm This 
semigroup is generated by the Hamiltonian $H$:
$$
\left|F(t)\rangle\langle F(t)\right|=e_+^{-iH^\times
t}\left|F(0)\rangle\langle F(0)\right|e_+^{iHt}
$$
It is {\it not} the semigroup
of quantum statistical mechanics of open systems generated by a}
%Liouvillian $L$, i.e., this is not the irreversible time evolution of
%open systems under external influences [17].}
%For the state $\rho$ of such open systems, one has in
%place of (2.8)
%$$
%{{\partial\rho(t)}\over{\partial
%t}}=L\rho(t)=-i[H,\rho(t)]+{\cal I}\rho(t),
%$$
%where $H$ is the Hamiltonian of the open system and ${\cal I}$
%is the interaction of the external reservoir upon the system, e.g.,
%$$
%{\cal I}=\sum_{\alpha=1,2,..}([V_{\alpha}\rho(t),V_{\alpha}^\dagger]+
%[V_{\alpha},\rho(t)V_{\alpha}^\dagger])
%$$
%\n The time evolution semigroup for open systems is
%$$
%\rho(t)=\Lambda(t)\rho(0),\quad {\rm where}\ \Lambda(t)=e^{Lt},\ t\geq0
%$$
%\n $\rho(t)$ describes the state of an {\it open} system acted upon by
%an external reservoir (environment, measurement apparatus, quantum
%reservoir, etc.)
%%%%%%%%%%%%%%%%%%%%%%%%%%%%%%%%%%%%%%%%%%%%%%%%%%%%%%%
%\footnote{}{\n For such open systems,
%$$
%{{\partial\rho(t)}\over{\partial
%t}}=L\rho(t)=-i[H,\rho(t)]+{\cal I}\rho(t)
%$$
%where $H$ is the Hamiltonian of the open system and ${\cal I}$
%is the interaction of the external reservoir upon the system, e.g.,
%$$
%{\cal I}=\sum_{\alpha=1,2,..}([V_{\alpha}\rho(t),V_{\alpha}^\dagger]+
%[V_{\alpha},\rho(t)V_{\alpha}^\dagger])
%$$
%\n The time evolution semigroup for open systems is
%$$
%\rho(t)=\Lambda(t)\rho(0),\quad {\rm where}\ \Lambda(t)=e^{Lt},\ t\geq0
%$$
%
%\n $\rho(t)$ describes the state of an {\it open} system acted upon by
%an external reservoir (environment, measurement apparatus, quantum
%reservoir, etc.)}
%%%%%%%%%%%%%%%%%%%%%%%%%%%%%%%%%%%%%%%%%%%%%%%%%%%%%%%%%%%%%%%%%%%%
$$
F(t)=U_+^\times(t)F_0^-=e_+^{-iH^\times t}F_0^-\leqno(2.15a)
$$
$$
{\rm for}\ t\geq0\ {\rm only.}\leqno(2.15b)
$$
\n If we use the quantum mechanical state operators with 
semigroup time evolution,
$$
W^G(t)\equiv|F(t)\rangle\langle
F(t)|=e^{-iH^\times t}W^G(t_0)e^{iHt},\qquad t\geq0,\leqno(2.16)
$$
to calculate the quantum mechanical probabilities, then for these
calculated probabilities we obtain 
$$\leqalignno{
P_{W^G(t)}(\Lambda_0)={\rm Tr}(\Lambda(t_0)W^G(t))&={\rm
Tr}(\Lambda(t)W^G(t_0))&(2.17a)\cr
&\phantom{000}t\geq t_0=0.&(2.17b)\cr}$$
This means that they fulfill the same conditions 
as the experimental
probabilities (2.12a), 
(2.12b) and (2.13).\footnote{}{\noindent 
Liouvillian $L$, i.e., this is not the 
irreversible time evolution of open systems under external influences [17].
For the state $\rho$ of such open systems, one has in
place of (2.8)
$$
{{\partial\rho(t)}\over{\partial
t}}=L\rho(t)=-i[H,\rho(t)]+{\cal I}\rho(t),
$$
where $H$ is the Hamiltonian of the open system and ${\cal I}$
is the interaction of the external reservoir upon the system, e.g.,
$$
{\cal I}=\sum_{\alpha=1,2,..}([V_{\alpha}\rho(t),V_{\alpha}^\dagger]+
[V_{\alpha},\rho(t)V_{\alpha}^\dagger])
$$
\n The time evolution semigroup for open systems is
$$
\rho(t)=\Lambda(t)\rho(0),\quad {\rm where}\ \Lambda(t)=e^{Lt},\ t\geq0
$$

\n $\rho(t)$ describes the state of an {\it open} system acted upon by
an external reservoir (environment, measurement apparatus, quantum
reservoir, etc.)}

In particular the probabilities
are not defined unless the preparation $\Rightarrow$ registration arrow
of time (2.17b) is fulfilled, because
the time evolution
$$W^G(t)=e^{-iH^\times t}W^G_0e^{iHt}\qquad {\rm or}\
(2.18a')\ \Lambda(t)=e^{iHt}\Lambda_0e^{-iH^\times t}\leqno(2.18a)
$$
is a semigroup evolution and only defined for
$$
t>t_0=0.\leqno(2.18b)
$$
The physical meaning of the initial time $t_0$ for a decaying system in the
state $W^{\rm G}$ will be
discussed in section~4 below. Mathematically, it is given by the
initial time $t=0$ of the Cauchy problem (2.14).

This semigroup-arrow of time (2.15b), (2.17b), (2.18b) is the formulation
in the mathematical theory of the experimental preparation $\Rightarrow$
registration arrow of time (2.12).\footnote{$^2$}
{Since the semigroup time
evolution (2.15) or (2.18)  is not possible
in the  Hilbert
space, i.e., $F_0^-\not\in{\cal H}$, 
people who wanted to retain the standard Hilbert space theory
but were aware of the quantum mechanical preparation $\Longrightarrow$
registration arrow of time
had to
extrapolate (2.18) to negative times, therewith
eliminating the experimental preparation $\Longrightarrow$ registration
arrow of time and causality
from the mathematical theory [18].} 

2.)  The universe, when considered as a quantum physical system, 
must also be in
a state $\rho$ (a pure state
$\rho=|\phi\rangle\langle\phi|$, or a mixture) with asymmetric
 time evolution [19].  Its arrow of time must be identical with
the traditional cosmological arrow of time and the time $t=t_0=0$, at which
the initial state of the universe
$\rho$ has been prepared, is the time of the big bang.

The general quantum mechanical (a priori) probabilities predicted 
for the
observable represented by the
projection operator $P^1_{\alpha_1}(t_1)$ (``yes-no observations'') 
are according
to (2.17),
$$\leqalignno{
{\cal P}(\alpha_1,t_1)\equiv{\cal P}_\rho(P^1_{\alpha_1}(t_1))&={\rm
Tr}(P^1_{\alpha_1}(t_1)\rho)={\rm
Tr}(P^1_{\alpha_1}(t_1)\rho P^1_{\alpha_1}(t_1))&(2.19a)\cr
\phantom{00000}{\rm for}\phantom{000}t_1>t_0&=0\ {\rm only}&(2.19b)\cr}
$$
The time ordering (2.19b) is the same as the semigroup arrow of time
(2.17b) in the quantum mechanics of measured
systems.  Applied to experiments performed on quantum systems in the
laboratory it leads to the preparation
$\Longrightarrow$ registration arrow of time (2.12b).  Like in the quantum
mechanics of measured systems, (2.19b)
is an expression of causality.

The quantum mechanical probabilities (2.19) of projection operators
$P^i_{\alpha_i}(t_i)$ can be generalized to
probabilities of histories [3,~20].

A history is a time ordered product of different
projection operators (labeled by $\alpha_i$) for
different observables (labeled by $i$):
$$
\leqalignno{
C_\alpha&=P_{\alpha_1}^1(t_1)...P_{\alpha_i}^i(t_i)...
P_{\alpha_n}^n(t_n);\quad  t_n>t_{n-1}>...>t_2>t_1.&(2.20)\cr
\noalign{\hbox{\rm with}}
P_{\alpha_i}^i(t_i)&=e^{iH(t_i-t_{i-1})}P_{\alpha_i}^i(t_{i-1})
e^{-iH(t_i-t_{i-1})};&(2.21a)\cr
&\phantom{000}t_i-t_{i-1}>0&(2.21b)\cr}
$$

This definition of histories is suggested by the following considerations:

\n Let $P^i_{\alpha_i}$ be the $\alpha_i$-th projector of (what we denote as)
the $i$-th observable
$A^i=\Sigma_\alpha a^i_\alpha P^i_\alpha$\qquad$i=1,\ 2,\ 3,\ \ldots$.
Then, starting with the operator $\rho=\rho(t_0)$ of (2.19a), 
one can define a sequence of effective density operators
$\rho^{\rm eff}(t_1),\cdots,\rho^{\rm eff}(t_{n-1})$,  
and  
one can  predict a sequence of
probabilities
${\cal P}(\alpha_2t_2;\alpha_1t_1),{\cal
P}(\alpha_3t_3;\alpha_2t_2;\alpha_1t_1)\ldots{\cal P}(\alpha_n
t_n;\cdots\alpha_1t_1)$. These density operators and probabilities are
listed below: 
$$\leqalignno{
\rho^{\rm eff}(t_1)&={{P^1_{\alpha_1}(t_1)\rho(t_0)P^1_{\alpha_1}(t_1)}\over{{\rm
Tr}(P^1_{\alpha_1}(t_1)\rho(t_0)P^1_{\alpha_1}(t_1))}}=N_1P^1_{\alpha_1}(t_1)\rho(t_0)P^1_{\alpha_1}(t_1)&(2.22a)\cr
& {\rm for}\ t_1>t_0\ {\rm only};&(2.22b)\cr}
$$
(the second equality in (2.22a) defines the normalization factor $N_1$)\hb
\n and
$$
{\cal P}(\alpha_2t_2;\alpha_1t_1)=N_1{\rm
Tr}(P^2_{\alpha_2}(t_2)\rho^{\rm eff}(t_1)P^2_{\alpha_2}(t_2))\leqno(2.23a)
$$
$${\rm for}\
t_2>t_1\ {\rm only.} \leqno(2.23b)
$$
Continuing in this way for $n=3,\ 4,\ \ldots$,

$$
\eqalign{
\rho^{\rm eff}(t_{n-1})&={{P^{n-1}_{\alpha_{n-1}}(t_{n-1})\rho^{\rm eff}(t_{n-2})P^{n-1}_
{\alpha_{n-1}}(t_{n-1})}\over{{\rm
Tr}(P^{n-1}_{\alpha_{n-1}}(t_{n-1})\rho^{\rm eff}(t_{n-2})P^{n-1}_
{\alpha_{n-1}}(t_{n-1}))}}\cr
&\cr
&=N_{n-1}P^{n-1}_{\alpha_{n-1}}(t_{n-1})
\rho^{\rm eff}(t_{n-2})P^{n-1}_{\alpha_{n-1}}(t_{n-1});\cr
&\cr
&={{P^{n-1}_{\alpha_{n-1}}(t_{n-1})\cdots
P^1_{\alpha_1}(t_1)\rho(t_0)P^1_{\alpha_1}(t_1)\cdots
P^{n-1}_{\alpha_{n-1}}(t_{n-1})}\over{{\rm
Tr}(P^{n-1}_{\alpha_{n-1}}(t_{n-1})\cdots
P^1_{\alpha_1}(t_1)\rho(t_0)P^1_{\alpha_1}(t_1)\cdots
P^{n-1}_{\alpha_{n-1}}(t_{n-1}))}}}\leqno(2.24a)
$$
$$
t_{n-1}>t_{n-2}>\cdots>t_1>t_0\leqno(2.24b)
$$
\n and
$$
{\cal P}(\alpha_nt_n;\cdots\alpha_1t_1)={{{\rm
Tr}(P^n_{\alpha_n}(t_n)\rho^{\rm eff}(t_{n-1}))}\over{{\rm
Tr}(P^{n-1}_{\alpha_{n-1}}(t_{n-1})\rho^{\rm
eff}(t_{n-2})P^{n-1}_{\alpha_{n-1}} 
(t_{n-1}))}};\leqno(2.25a)
$$
$$
t_n>t_{n-1}\leqno(2.25b)
$$
\n or
$$
{\cal P}(\alpha_n,t_n;\dots;\alpha_1,t_1)=N_{n-1}{\rm
Tr}(P^n_{\alpha_n}(t_n)\dots P^1_{\alpha_1}(t_1)\rho(t_0)P^1_{\alpha_1}(t_1)
\dots P^n_{\alpha_n}(t_n));\leqno(2.26a)
$$
$$
{\rm for}\ t_n>\ldots>t_0\ {\rm only}\leqno(2.26b)
$$

The time ordering or arrow of time (2.23b)$\ldots$ (2.26b) is again
the same as the semigroup arrow of
time (2.18b), and (2.21) is the same as the semigroup evolution ($2.18'$)
(in the
Heisenberg picture)
for the
observable $\Lambda$  in the quantum theory of measured systems.

The probability (2.25a), (2.26a) is the probability of the history defined
in (2.20)
$$
{\cal P}(\alpha_nt_n\ldots\alpha_1t_1)=N_n{\rm
Tr}(C_\alpha\rho(t_0)C_\alpha)\leqno(2.27)
$$
One can consider alternative projection operators
$$C_\alpha'=P^{1^\prime}_{\alpha^{\prime}_1}(t_1)
P^{2^{\prime}}_{\alpha_{2'}}(t_2)\ldots
P^{n'}_{\alpha^{\prime}_n}(t_n)\leqno(2.28a)
$$
but a physical meaning can only be given to these products for the time
ordering
$$t_n>t_{n-1}...>t_1.\leqno(2.28b)$$

This time ordering, identical with the time ordering
(2.22b)$\ldots$(2.24b), is a calculational consequence of
the restriction (2.21b) postulated [3,~19] for the time 
evolution of the projectors.
The restricted time evolution (2.21) is a semigroup evolution
generated by the Hamiltonian of the
closed quantum system.
Obviously the semigroup (2.21), (2.18) and (2.16)
is the same semigroup applied to
different observables, 
$P^i_{\alpha_i}$ and $\Lambda$ 
respectively of different quantum systems, namely
the quantum universe and the
quasi-stable particle.  The semigroup character of the time evolution
(2.18$^{\prime}$) --- or of (2.18)
in the Schroedinger picture"---was inferred from restrictions imposed by
observational limitations in a
laboratory experiment with quantum systems, namely from the preparation
$\Longrightarrow$ registration arrow
of time.  The semigroup character of the time evolution (2.21) and the
time ordering (2.28b) were postulated 
for the quantum universe because of the special initial state
associated to the big bang [19].
From the way the time ordering appears in the probabilities for the
laboratory experiments (2.17) and in the probabilities of the
histories $(2.19)\ldots(2.25)$, it is clear that 
both time orderings 
express the same
arrow of time. If our universe is a closed quantum system as suggested
by [3], the semigroup arrow for the resonances is subsumed under the
cosmological arrow of time, or vice versa. This arrow of time
``may {\it not} be attributed to the thermodynamic arrow of
an external measuring apparatus (for the laboratory experiment) 
or larger
universe'' (for the quantum
universe). It is a
``fundamental quantum mechanical distinction between the past
and future'' [3]. 

As mentioned above, a semigroup evolution that could give a theoretical
description of this arrow of
time is impossible in the standard Hilbert space quantum
mechanics. Therefore, in order to make the semigroup postulate (2.21)
possible and to allow for a semigroup solution (2.15) of the quantum
mechanical Cauchy problem, one must develop a new mathematics. We shall
present the mathematics that is capable of a time asymmetric quantum 
theory in
the following section.

%%%%%%%%%%%%%%%%%%%%%%%%%%%%%%%%%%%%%%%%%%%%%%%%%%%%%%%%%%%%%%%%%%%%%%%%%%%%%%%
%The postulate (2.21) also avoids another inconsistency, if one wants to
%interpret  $\rho^{\rm eff}(t_1)$ as the state
%resulting from a measurement of the observable $P_{\alpha_1}$, on the state
%$\rho(t_0)$.  The von Neumann axiom of
%ideal measurement of the first kind (collapse of the wave function)
%states that 
%$\rho^{\rm after}(t_0)=P_{\alpha_1}\rho(t_0)P_{\alpha_1}$. This means
%that the ``collapse'' is supposed to happen instantaneously, but realistically
%every measurement takes time.   One avoids this
%inconsistency by using the time evolution (2.21) which defines a
%sequence of ``new initial'' states $\rho^{\rm
%eff}(t-1),\cdots\rho^{\rm eff}(t_n),$ where
%$t_1-t_0,\cdots,t_n-t_{n-1}$ are positive, finite time intervals
%during which a (more) realistic measurement can take place.
%%%%%%%%%%%%%%%%%%%%%%%%%%%%%%%%%%%%%%%%%%%%%%%%%%%%%%%%%%%%%%%%%%%%%%%%%%%%%%%

\vfill\eject
\noindent{\bf 3. A Mathematical Theory for Time Asymmetric Quantum Physics}

\n Our empirical consideration in section~2 has led us to the
postulate of a time
evolution semigroup (2.21) or (2.18).  Here
we want to discuss a mathematical theory of
quantum physics  for which a semigroup evolution exists. 
%%%%%%%%%%%%%%%%%%%%%%%%%%%%%%%%%%%%%%%%%%%%%%%%%%%%%%%%%%%%%%%%%%%%%%%%%%%%%%%%%%%%%%%%%%% 
%As already mentioned,
%in standard quantum mechanics the symmetry
%transformations (e.g., Galileo transformations, Poincare
%transformations) are described by a unitary group
%representation in $\cal H$. In particular, the time evolution is
%unitary and reversible, and it is given by
%$U^\dagger(t)={\rm exp}(-iHt),-\infty<t<\infty,\  {\rm of}\
%(2.9)\ldots$ (2.11).
%Though there seems to be no empirical reason that forbids time
%asymmetric quantum mechanics, the widespread conclusion from this was that
%irreversible time evolution of isolated quantum mechanical
%systems is impossible.
%%%%%%%%%%%%%%%%%%%%%%%%%%%%%%%%%%%%%%%%%%%%%%%%%%%%%%%%%%%%%%%%%%%%%%%%%%%%%%%%%%%%%%%%%%%%%%%

In a linear space with a scalar product $\Phi_{\rm alg}$, which we need for
the calculational rules of
quantum mechanics, the simplest modification
that allows Hamiltonian generated
semigroups is to choose instead of the Hilbert space topology 
a locally convex topology.  If one also wants the
Dirac formalism (i.e., kets, the basis
vector expansion (2.5) etc.), then one has to choose a Rigged Hilbert Space
(RHS) or Gelfand triplet.
$$
\Phi\subset{\cal H}\subset\Phi^\times\leqno(3.1)
$$
The triplet of spaces in a Rigged Hilbert Space
 $\Phi\subset{\cal H}\subset~\Phi^\times$
results from  three different topological completions of the same
algebraic (pre-Hilbert) space $\Phi_{\rm alg}$ of section~2, [21].
Completion means adjoining to $\Phi_{\rm alg}$ the (limit elements of)
convergent
(Cauchy) sequences with respect to a topology.
The completion of $\Phi_{\rm alg}$
with respect to the norm
$\|\varphi\|=\sqrt{(\varphi,\varphi)},\ \varphi\in\Phi_{\rm
alg}$
is the Hilbert space ${\cal H}$.
The topology or meaning of convergence defined by the norms we denote by
${\cal T}_{\cal H}$.
The completion of $\Phi_{\rm alg}$ with
respect to a finer locally convex, nuclear topology, which we denote by
${\cal T}_\Phi$ (and which is usually given by a
countable
number of norms [21]), is denoted by $\Phi$.
Then one has $\Phi_{\rm alg}\subset\Phi\subset{\cal H}$ (because $\Phi$ and
${\cal H}$ contain all elements of $\Phi_{\rm alg}$ plus the limit
elements of Cauchy
sequences in $\Phi_{{\rm alg}}$), and $\Phi\subset{\cal H}$
holds because ${\cal T}_\Phi$ is chosen to be finer or stronger than
${\cal T}_{\cal H}$ and there are consequently 
more ${\cal T}_{\cal H}$ Cauchy sequences than ${\cal T}_{\Phi}$ 
Cauchy sequences.
We also consider the space of
${\cal T}_{\cal H}$-~continuous and of ${\cal T}_{\Phi}$-~continuous
functionals.  
${\cal H}^\times$ is the space of ${\cal T}_{\cal H}$ continuous antilinear
functionals $\psi$
 on the space ${\cal H}:\ \psi:\ \phi\in{\cal H}\rightarrow\psi(\phi)
\in{I\kern-.55em C}$, and ${\cal H}={\cal H}^\times,\
\psi(\phi)=(\phi,\psi),$ by a mathematical theorem.
$\Phi^\times$ is the space of ${\cal T}_\Phi$-continuous, antilinear
functionals $F$ on the space $\Phi:\
F\ :\ \phi\in\Phi\rightarrow\
F(\phi)\equiv\langle\phi|F\rangle
\in{I\kern-.55em C}.$ One has ${\cal H}^\times\subset\Phi^\times$ and
the bra-ket $<|>$ becomes an extension of the scalar product.  Thus
one obtains the Gel'fand
triplet (3.1).

Dirac kets are elements of $\Phi^\times$, but there are also other
$|F\rangle\in\Phi^\times$ besides the Dirac kets.  Dirac's algebra of
observables is an algebra of
continuous operators in $\Phi$ (observables cannot be continuous
operators in ${\cal H}$).

\n For a ${\cal T}_\Phi$-continuous linear operator $A$,
its conjugate operator $A^\times$ is defined by
$$
\langle A\phi|F\rangle=\langle\phi|A^\times|F\rangle,\
\forall\phi\in\Phi\ {\rm and}\ \forall F\in\Phi^\times\leqno(3.2)
$$
$A^\times$ is a continuous operator in $\Phi^\times$.
Then for each observable $A$, one has a triplet of operators
$$
A^{\dagger}\big|_{\Phi}\subset A^\dagger\subset A^\times\leqno(3.3)
$$
where $A^{\dagger}$ is the Hilbert space adjoint operator of $A$ and
$A^{\dagger}|_\Phi$ is its restriction to
the subspace $\Phi$.
Generalized eigenvectors are defined for continuous operators.
A vector $|F\rangle\in\Phi^\times$ is a generalized eigenvector of
the ${\cal T}_\Phi$-continuous operator $A$ if for some complex number
$\omega$ and for all $\phi\in\Phi$,
$$
\langle A\phi|F\rangle=\langle\phi|A^\times|F\rangle
=\omega\langle \phi|F\rangle\leqno(3.4)
$$
This is also written as 
$$
A^\times|F\rangle=\omega|F\rangle\leqno(3.5)
$$
(or, even as
$A|F\rangle=\omega|F\rangle$
if $A^\dagger$ is a self-adjoint operator).

If $A$ is the (self adjoint) 
Hamiltonian $H$ of a quantum physical system, then
$\Phi^\times$ contains the
Dirac kets
$$
H^\times|E^-\rangle=E|E^-\rangle,\quad E\geq0\leqno(3.6)
$$
$\Phi^\times$ can also contain
generalized eigenvectors with complex eigenvalues, as e.g.,
$$
H^\times\big|E_{\rm R}-i\Gamma/2^-\rangle=
\left(E_{\rm R}-i\Gamma/2\right)\big|E_{\rm R}-i\Gamma/2^-\rangle,\leqno(3.7)
$$
which we call Gamow vectors or Gamow kets [22].

There is not only one space $\Phi$, but there are many
(locally convex, nuclear,
countably normed) topologies ${\cal T}_\Phi$, which lead to different
completions $\Phi$
of $\Phi_{\rm alg}$ (with the same ${\cal H}$).
The choice of $\Phi$ depends on the particular physical problem at hand,
e.g., $\Phi$ can be chosen such that the algebra of observables of a
particular physical system is an
algebra of ${\cal T}_\Phi$ continuous operators.

Further,
in section~2 we said that we need to distinguish meticulously between
states and observables.
In order to be able to also distinguish mathematically between states and
observables
we have to introduce one space for states, which we call $\Phi_-$, and
another  space for observables,
which we call $\Phi_+$.  In general $\Phi_+\not=\Phi_-$, but
$\Phi_+\cap\Phi_-\not=\{0\}$. 
The state prepared by the preparation apparatus (e.g., accelerator) we
denote by $\phi^+$, thus
$\phi^+\in\Phi_-$.  The observable registered by the registration apparatus
(e.g., detector) we denote by
$|\psi^-\rangle\langle\psi^-|$, thus
$\psi^-\in\Phi_+$, (cf.~Appendix B for the scattering experiment).
Therefore we need two Rigged Hilbert Spaces, one for
prepared in-states $\phi^+$:
$$
\phi^+\in\Phi_-\subset{\cal H}\subset\Phi_-^\times,\leqno(3.8)
$$
and the other for the registered observables
$|\psi^-\rangle\langle\psi^-|$ or
detected out-states $\psi^-$:
$$
\psi^-\in\Phi_+\subset{\cal H}\subset\Phi_+^\times\leqno(3.9)
$$
In here the space ${\cal H}$ is the same Hilbert space (with the same
physical interpretation).

Mathematically one can define the spaces of the vectors $\Phi$ by the spaces of
their energy wave functions $\langle E|\phi\rangle$\footnote{$^1$}{In
the same way as one can define the Hilbert space $\cal H$ by the space
of Lebesgue square integrable functions ${\cal H}\ni h\Leftrightarrow
h(E)\in L^2[0,\infty)$, where the functions $h(E)$ are uniquely
determined only up to a set of Lebesgue measure zero, which
is a complicated and unphysical notion, cf.~section~$3$ ref.~[26]}:
$$
\phi^+\in\Phi_-\ \Leftrightarrow\ \langle^+E|\phi^+\rangle\in{\cal
S}\cap{\cal H}_-^2|_{{\IR}^+},\ ({\rm well\ behaved\ Hardy\ functions\ in}\
{I\kern-.55em C}^-).\leqno(3.10)
$$
$$
\psi^-\in\Phi_+\ \Leftrightarrow\ \langle^-E|\psi^-\rangle\in{\cal
S}\cap{\cal H}_+^2|_{{\IR}^+},\ ({\rm well\ behaved\ Hardy\ functions\ in}\
{I\kern-.55em C}^+).\leqno(3.11)
$$
The notation in here is the following:  ${I\kern-.55em
C}^+({I\kern-.55em C}^-)$ denotes
the open upper
(lower) half of the complex energy plane of the second Riemann sheet for the
analytically continued $S$-matrix,
and ${\cal H}^2_{\mp}$ denotes the Hardy
class functions [23] and ${\cal S}$ the Schwartz space functions.
This explains the notation $\Phi_-$ and $\Phi_+$ for the spaces. 
The subscript refers to the subscript in 
the standard notation of mathematics for Hardy class functions
(${\cal H}^p_-,{\cal H}^p_+$ respectively).
The superscripts for $\phi^+$ (in-states) and $\psi^-$ (out-states)
are
the most common convention in scattering theory, cf.~Appendix B.

Thus, in the physical interpretation, for each species of quantum physical
system
one has a pair of RHS's, (3.8) and (3.9).  Whereas the ``in-state''
$\phi^+\in\Phi_-$
describes the state that is physically defined by the preparation apparatus, 
the
``out-state'' $\psi^-\in\Phi_+$ describes the observable that is physically
defined by
the registration apparatus.

It is by this clear differentiation between the set of vectors $\{\phi^+\}$
which are
admitted as in-states and the set of vectors $\{\psi^-\}$ which are admitted as
out-observables that the RHS-theory differs from the usual scattering
theory,
where $\{\phi^+\}=\{\psi^-\}=\Phi\subset{\cal H}$ (cf.~the asymptotic
completeness condition according to which
$\{\phi^{+}\}=\{\phi^{-}\}={\cal H}$).  According to (3.10) and (3.11), 
$\Phi_-$ and $\Phi_+$ are different dense
subspaces of the same Hilbert space ${\cal H}$ (which are both complete
with respect
to a stronger topology than ${\cal T}_{\cal H}$) with
$$
\Phi_+\cap\Phi_-\not=\{0\},\ {\rm and}\ \Phi=
\Phi_++\Phi_-\  {\rm is\  also\  dense\  in}\  {\cal H}.\leqno(3.12)
$$
After the RHS's (3.8) and (3.9) have been chosen to be the Hardy class
spaces (3.10)
and (3.11), the semigroup of section 2 turns up naturally from the
mathematics.
How one could empirically conjecture the RHS's of Hardy class will not be
discussed
here [24].

To obtain the semigroups we start with the unitary group of time evolutions in
the Hilbert space ${\cal H}$.
$$
U(t)=e^{iHt},\qquad\qquad U^\dagger(t)=e^{-iHt}\leqno(3.13)
$$
where $U^\dagger(t)$ denotes the Hilbert space 
adjoint of $U(t).$\footnote{$^2$}
{Note that in ${\cal H}$ the right hand side of (3.13) 
is not defined by the exponential series
$$
I+{iHt\over1!}+{(iH)^2t^2\!\over2!}+{(iH)^3t^3\over 3!}+\ldots
$$
which only converges with respect to ${\cal T}_{\cal H}$ on a dense subspace of
analytic vectors in ${\cal H}$, but
by the Stone- von Neumann calculus.}

We first turn to the RHS (3.9) and consider
$$
U_+(t)\equiv U(t)|_{\Phi_+}\subset U(t),\ \ {\rm and}\
U(t)^\dagger\subset U^\times_+(t)\leqno(3.14)
$$
It can be shown that, as a consequence of the mathematical properties
of $\Phi_+$, the restriction of $U(t)$ to $\Phi_+,\
U_+(t)$, is a ${\cal T}_{\Phi_+}$-continuous 
operator only for $0\leq t<\infty$.
Therefore its conjugate operator 
$U_+^\times(t)$, which is an extension of the Hilbert space adjoint operator
$U^\dagger(t)$, is well defined (by (3.2)) and
continuous
for
$0\leq t\leq\infty$ only.
Thus in $\Phi^\times_+$ we have only the semigroup
$$
U_+^\times(t)=(e^{iHt}|_{\Phi_+})^\times\equiv e^{-iH^\times t}_+,\qquad
0\leq t<\infty\leqno(3.15)
$$

The same considerations apply to the other RHS (3.8).  One considers\break\hfil
$U_-(t)\equiv U(t)|_{\Phi_-}\subset U(t)$, and its  conjugate
$U(t)^\dagger\subset U^\times_-(t)$,
and proves mathematically that
$U_-(t)$ is a ${\cal T}_{\Phi_-}$-continuous  operator  only  for
$-\infty\leq t\leq0.$ Therefore
$U^\times_-(t)$ is defined and continuous for $-\infty\leq t\leq0$ only and
one has in $\Phi^\times_-$ 
the semigroup
$$
U_-^\times(t)=\left(e^{iHt}\big|_{\Phi_-}\right)^\times\equiv e_-^{-iH^\times
t},\ -\infty<t\leq0\leqno(3.16)
$$
Thus in the RHS (3.8) for the prepared states one has the semigroup (3.16)
for times $t\leq t_0=0$, and in the RHS (3.9) for the registered
observables one has the semigroup
(3.15) for times $t\geq t_0=0$. Since $t=t _0=0$ is the time by which the state
has been
prepared and the registration of the observable can begin, this separation
of the
(mathematical) group (3.13) into the two semigroups (3.16) and (3.15)
reflects the
situation envisioned on empirical grounds in section 2.  The
scattering (e.g., 
resonance scattering) process is separated into two parts, the preparation part
dealing with the preparation of the state $\phi^+\in\Phi$ and the
registration part dealing with the registration of the observable (or
detection of
the out-state) $\psi^-\in\Phi_+$.  The time $t=0(t_0)$ is the time at which the
preparation is completed and the registration can commence; the meaning
of $t_0$
will be discussed in detail in section~4.

In addition to the vectors $\phi^+$ and $\psi^-$ defined by the
apparatuses, there also are the
vectors in
$\Phi^\times_\pm$ which are outside of ${\cal H}$:
$$
|E,\theta_{\rm p},\varphi_{\rm p}^\mp\rangle\in\Phi_\pm^\times,\quad
({\rm Dirac's\  scattering\  states}),\leqno(3.17)
$$
where   $(\theta_p,\varphi_p)$ denotes 
the direction  of momentum;\hb
and the
$$
\psi^{\rm G}=|E_{\rm R}\mp
i{\Gamma/2},j,j_3^\mp\rangle\in\Phi_{\pm}^\times\qquad 
({\rm Gamow's\  resonance\  states})\leqno(3.18)
$$
with the property
$$
H^\times|E_{\rm R}-i{{\Gamma}/{2}}^-\rangle=
\left(E_{\rm R}-i\Gamma/2\right)|E_{\rm
R}-i{{\Gamma}/{2}}^-\rangle.\leqno(3.19)
$$
%%%%%%%%%%%%%%%%%%%%%%%%%%%%%%%%%%%%%%%%%%%%%%%%%%%%%%%%%%%%%%%%%%%%%%%%%%%%%%%%%%%%%%%%%%
%Dirac's scattering state vectors are copiously used, though their mathematical
%meaning is usually not fully appreciated.
%Gamow's resonance state vectors are in great disrepute in spite of the fact,
%that their origin goes back about the same long time [15].
%%%%%%%%%%%%%%%%%%%%%%%%%%%%%%%%%%%%%%%%%%%%%%%%%%%%%%%%%%%%%%%%%%%%%%%%%%%%%%%%%%%%%%%%%%%

In the RHS theory Dirac kets and Gamow vectors are mathematically very
similar.  Both are generalized
eigenvectors of self-adjoint Hamiltonians in the sense of (3.2) and are
equally well defined, (though
the choice of spaces $\Phi$ for which Gamow kets can be defined is smaller
than for Dirac kets since the
former also requires some analyticity properties as for Hardy class spaces
$\Phi_+$).  Dirac kets and Gamow kets just differ
in their eigenvalues; whereas Dirac's scattering state vectors in 
$\Phi_{+}^{\times}$ or
$\Phi_{-}^{\times}$ have real (except for
the $\pm i0$) eigenvalues corresponding to the scattering energies, Gamow
kets  have complex
eigenvalues corresponding to the resonance pole of the $S$-matrix (see below).

The Gamow vectors $\psi^{\rm
G}=|E_R-i\Gamma/2\rangle\sqrt{2\pi\Gamma}\in\Phi_+^\times$ 
have a semigroup time evolution and obey an exponential law:
$$
\psi^{\rm G}(t)\equiv U_+^{\times}(t)\psi^{\rm G}=
e^{-iE_{\rm R}t}e^{-\Gamma t/2}|E_{\rm R}-i\Gamma/2^-\rangle,\qquad 
t\geq0\leqno(3.20)
$$
This is a formal consequence of applying the right hand side 
of (3.15) to $\psi^{\rm
G}$ and using (3.19). But for
the mathematical proof of (3.20), in particular of the semigroup character,
the whole mathematical
apparatus of the RHS of Hardy class is needed [21].

There are other Gamow vectors $\tilde\psi^G=|E_{\rm
R}+i\Gamma/2^+\rangle\sqrt{2\pi\Gamma}\in\Phi^\times_-$, and there is
another semigroup (3.16), $e^{-iH^\times t}_-$ for
$t\leq0$ in $\Phi_-\subset{\cal H}\subset\Phi^\times_-$ with the asymmetric
evolution
$$
\tilde\psi^G(t)=e^{-iH^\times t}|E_{\rm R}+i\Gamma/2^+\rangle=
e^{-iE_{\rm R}t}e^{\Gamma t/2}|E_{\rm R}+i\Gamma/2^+\rangle,\ t\leq0
\leqno(3.21)
$$
Gamow vectors have the following features:
\item{1.}
They are derived as functionals of the resonance pole term at $z_{\rm
R}=E_{\rm R}-i\Gamma/2$ (and at
$z^*_{\rm R}=E_{\rm R}+i\Gamma/2)$ in the second sheet  of the analytically
continued $S$-matrix [9,~25].
\item{2.}
They have a Breit-Wigner energy distribution
$|\langle^{-}E|\psi^{G}\rangle|^{2}={{\Gamma}\over{2\pi}}
{{1}\over{(E-E_{R})^{2}+\left({{\Gamma}\over{2}}\right)^{2}}}
\rightarrow \delta (E-E_{R})$ for ${{\Gamma}\over{E_{R}}}\rightarrow 0$
which extends to negative energy values on the second sheet 
indicated in the representation
$$
|\psi^{G}\rangle=i\sqrt{{{\Gamma}\over{2\pi}}}
\int_{-\infty_{II}}^{+\infty}dE {{|E^{-}\rangle}\over
{E-\left(E_{R}-i\Gamma/2\right)}}\leqno(3.22)
$$
by $-\infty_{II}$ [9].
\item{3.}
The decay probability ${\cal P}(t)={\rm Tr}(\Lambda|\psi^{\rm
G}\rangle\langle\psi^{\rm G}|)$ of
$\psi^{\rm G}(t),\ t\geq0$, into the final non-interacting decay products
described by $\Lambda$ can be
calculated as a function of time, and from this the decay rate
$R(t)={d{\cal P}(t)\over dt}$ is obtained
by differentiation [26].
This leads to an exact Golden Rule (with the natural line width given by
the Breit-Wigner) and the
exponential decay law
$$
R(t)=e^{-i\Gamma t}\Gamma_\Lambda\qquad t\geq0
\leqno(3.23)
$$
where $\Gamma_\Lambda$ is the partial width for the decay products
$\Lambda$ ($\Gamma_\Lambda$=branching
ratio$\times\Gamma$). In the Born approximation ($\psi^{\rm
G}\rightarrow f^{\rm D}$, an eigenvector of $H_0=H-V;\
\Gamma/E_R\rightarrow0;\ E_R\rightarrow E_0$) this exact Golden Rule
goes into Fermi's Golden Rule No.~2 of Dirac.
\item{4.}
The Gamow vectors $\psi^{\rm G}_i$ are members of a ``complex'' basis
vector expansion [25]. In place of the well known
Dirac basis system expansion (Nuclear Spectral Theorem
of the RHS) given by
$$
\phi^+=\sum_n|E_n)(E_n|\phi^+)+ 
\int_0^{+\infty}dE|E^+\rangle\langle^+E|\phi^+\rangle\leqno(3.24)
$$
(where the discrete sum is over bound states, which we henceforth
ignore),
every prepared state vector $\phi^+\in\Phi_-$ can be expanded as
$$
\phi^+=\sum_{i=1}^N|\psi^{\rm G}_i\rangle\langle\psi^{\rm G}_i
|\phi^+\rangle+\int_0^{-\infty_{II}}dE|E^+\rangle\langle^+E|\phi^+\rangle
\leqno(3.25) 
$$
(where $-\infty_{II}$ indicates that the integration along the
negative real axis or other contour including around cuts is in the
second Riemann sheet of the $S$-matrix). $N$ is the number of
resonances in the system (partial wave), each one occurring  
at the pole position $z_{{\rm R}_i}=E_{{\rm R}_i}-i\Gamma_i/2$. This
allows us to mathematically isolate the exponentially decaying states
$\psi^{\rm G}_i$. 

The ``complex'' basis system expansion is rigorous. The
Weisskopf--Wigner approximate methods are tantamount to omitting the
background integral, i.e.,
$$
\phi^+{\mathop=^{\rm W-W}}\sum_{i=1}^N|\psi^{\rm G}_i\rangle
c_i\leqno(3.26)
$$
For instance, for the $K_L-K_S$ system with $N=2$,
$$
\phi^+=\psi^{\rm G}_Sb_S+\psi^{\rm G}_Lb_L
$$

The properties (3.18)--(3.25) are not independently postulated
conditions for the Gamow vectors but derived from each
other in the mathematical theory of the RHS. One can start for 
instance with the most widely accepted definition of the 
resonance by the pair of poles of the S-matrix (B.3)
at $z_{R}=E_{R}\mp i\Gamma/2$ and associate to it the 
Lippmann-Schwinger-Dirac ket $|z_{R}^{\mp}\rangle$ obtained
from analytic continuation in (B.3). Then one obtains the
Breit-Wigner energy distribution (3.22) from the Hardy class 
property (3.11) and vice versa. From (3.22), using (3.11)--
in particular the property of the Schwartz space $\cal S$--
one derives (3.19) as generalized eigenvalue equation
$\langle \psi^{-}|H^{\times\, n}|z_{R}^{-}\rangle
=z_{R}^{n}\langle \psi^{-}|z_{R}^{-}\rangle$, not only
for $n=1$ but for all powers $n$. The generalized eigenvalue equation 
(3.20) is also derived from the representation (3.22) but only
for $t\geq 0$ because of the Hardy class property (which in turn
was needed to justify the Breit-Wigner energy distribution 
for the pole term of the S-matrix). The Dirac basis vector
expansion (3.24) is fulfilled for every RHS, e.g., when
$\Phi$ is realized just by $\cal S$. The basis vector expansion
(3.25) follows by analytic continuation and therefore
requires the Hardy class property (3.10)(3.11). The
derivation of the exact Golden Rule [26] uses in addition the
Lippmann-Schwinger equation (B.4).
\vfill\eject
\noindent{\bf 4.  The Physical Interpretation and the Meaning of the
Initial Time}

\n The semigroup time evolution introduces a new
concept --- the time $t_0(=0)$ at which the
preparation of the state is completed and the
registration of the observable can
begin.
This is the most difficult new concept, because one is unprepared for it by
the school of thinking based
on the old time symmetric quantum mechanics.   For the state of our
universe as a whole, considered as a
{\it closed} quantum mechanical system,
there is no problem, because
we deal only with one single system and the time
$t_0$ is
the time of the creation of this single universe (big bag time).
Alternatively, we could 
consider this universe as a member of an ensemble of universes,
of which we
have access to only our universe.  Then the probabilities (2.19a)-
(2.26a)  are
the statistical probabilities (``relative frequencies'') of this
ensemble and we have 
the usual interpretation of quantum mechanics, where the density operator
$\rho$, $W$, the state vector $\phi^+$ or wave function
$\langle^+E|\phi^+\rangle$ is the mathematical
representative of an ensemble of microsystems. 

For an experiment performed on a quantum system in the laboratory,
the states prepared by a macroscopic preparation apparatus, i.e., states
described by
$\phi^+\in\Phi_-$ or $W=\sum_i
w_i|\phi^+_i\rangle\langle\phi^+_i|,\ \phi^+_i\in\Phi_-$, are best
interpreted as ensembles (e.g., the proton or electron beam prepared by an
accelerator).  But there are
other ``states" which are prepared by a macroscopic apparatus in
conjunction with a quantum scattering
process (e.g., resonance scattering), which are best interpreted as states
of a single microsystem.  For
their description the RHS offers, e.g., the Dirac kets (3.17) or the Gamow
kets (3.18).  From the basis vector expansion (3.25), we know that,
mathematically, the apparatus-prepared state $\phi^+$ can be
represented as the sum of a Gamow vector $\psi^{\rm G}$ and a
background integral. We shall now argue that the Gamow state can also
be isolated experimentally and discuss its creation time $t_0$ and
its asymmetric development in time.\footnote{$^1$}{For another
discussion of the impossibility of time reversing the development of a
decaying microphysical system, see T.~D.~Lee [16].} This  microphysical
irreversibility is the analogue of the arrow of time for
the state of our universe.

The best example is the decaying
state of the 
neutral Kaon system
because it is a wonderfully closed system, isolated from most external
influences (including the
electromagnetic field) whose (exact) evolution  in time is
probably entirely due to the
Hamiltonian of the neutral Kaon system and  free of external influences
like those mentioned in footnote 1 of section~2.
Since we are here only interested in the
fundamental concepts of decay,
we discuss
a simplified $K^{\circ}$-system for which the $K^{\circ}_L$ as well as
the CP violation is ignored [27].
%%%%%%%%%%%%%%%%%%%%%%%%%%%%%%%%%%%%%%%%%%%%%%%%%
%\footnote{$^5$}{For a
%detailed analysis of the $K_L-K_S$ system in the framework of time asymmetric
%quantum mechanics with and without CP violation, see A.~Bohm [18].}.
%%%%%%%%%%%%%%%%%%%%%%%%%%%%%%%%%%%%%%%%%%%%%%%%%%%%%%%%%%%

The process (idealized, because in the real experiment one does not use a
$\pi$---but a proton beam) by
which the neutral Kaon state is prepared is:
$$
\pi^-p\Longrightarrow\Lambda K^{\circ};\qquad\qquad
K^{\circ}\Longrightarrow\pi^+\pi^-.\leqno(4.1)
$$
$K^{\circ}$ is strongly produced with a time scale of
$10^{-23}\sec$.~and  it decays weakly, with a time scale of
$10^{-10}\sec,$ which is roughly the lifetime of the $K^\circ_S,\ \tau_{K_s}$.
Thus $t_0$, the time at which the preparation of the
$K^{\circ}$- state, which we call $W^{K^{\circ}},$ 
is completed and the registration can begin, is
very well defined.  (Theoretical uncertainty is $10^{-13}\tau_K$). 
A schematic diagram of a real experiment [28] 
is shown in Figure~1. The state
$W^{K^{\circ}}$ is
created instantly at the baryon target $T$ (and the baryon $B$ is
excited from the
ground state (proton) into the $\Lambda$ state, 
with which we are no further
concerned).  We imagine that a single particle $K^{\circ}$
is moving into the forward
beam direction, because somewhere at a distance, say at $d_2$ from
$T$, we
``see'' a
decay vertex
for  $\pi^+\pi^-$, i.e., a detector (registration apparatus) has been built
such that
it counts $\pi^+\pi^-$ pairs which are coming from the position $d_2$.  The
observable registered by the detector is the projection operator
$$
\Lambda(t_2)=\left|\pi^+\pi^-,t_2\rangle\langle\pi^+\pi^-,t_2\right|
=|\psi^{\rm out}(t_2)\rangle\langle\psi^{\rm out}(t_2)|\leqno(4.2)
$$
for those  $\pi^+\pi^-$ which originate from the fairly well
specified location $d_2$.  From the position (in the lab
frame) $d^{\rm lab}_2$, the four-momentum
$p$ of
the $K^{\circ}(=$ the $z$ component of the momentum of the $\pi^+\pi^-$
system) and
the mass $m_K$ of $K^{\circ}$, one obtains the time $t_2^{\rm rest}$
(in the $K^{\circ}$ rest frame) which the $K^{\circ}$
has taken to move from $T$ to $d^{\rm lab}_2$.  This
is given by the simple formula of relativity
$d^{\rm lab}_2=t^{\rm rest}_2{p\over m_K}$ which we write
$d_2=t_2{p\over m_K}$.

We do not have to focus at only one location $d_2$ but can count decay
vertices at
any distance $d$ (of the right order of magnitude).
The detector (described by the
projection operator
$\Lambda(t)\equiv\left|\pi^+\pi^-,t\rangle\langle\pi^+\pi^-,t\right|$
counts 
 the $\pi^+\pi^-$ decays 
at different times \hskip 1.5in $t=t_1,\ t_2,\ t_3,\dots$
(in the rest frame of the $K^\circ$), and these correspond to the
distances from the target $d_1=pt_1/m_{K},\
d_2=pt_2/m_{K}, \dots$ (in the lab frame).

One ``sees" the decay vertex $d_i$ for each single decay and imagines
a single decaying $K^\circ$ micro-system that had been
created on the target $T$ at time $t_0=0$ and then traveled
a time $t_i$ until it decayed at the vertex $d_i$.
We give the following interpretation to these observations:  a single
microphysical
decaying system $K^{\circ}$ described by $W^{K^{\circ}}$ has been
produced by a
macroscopic registration apparatus and a quantum scattering process, at a time
$t=0$.  Each count of the detector is the result of
the decay
of such a single microsystem.  This particular microsystem has lived for a
time $t_i$---the time that
it took
the decaying system to travel from the scattering center $T$ to the decay
vertex $d_i$.
The whole detector registers the counting rate
${\Delta N(t)\over{\Delta t}}\approx NR(t)$ as a function of
$d_i$, i.e., of
$t_i={{m_{k}}\over{p}}d_i$, for $\cdots t_i>\cdots t_2>t_1>t_0=0.$ 
($N$ is the total number of counts).

The counting rate ${\Delta N(t_i)\over\Delta t}$ is plotted as a function
of time $t$ (in the $K^{\circ}$
rest frame), in~Figure~2.

No $\pi^+\pi^-$ are registered for $t<0$, i.e., clicks of the counter for
$\pi^+\pi^-$ that would point
to a decay vertex at the position $d_{-1}$ in front of the target $T$ are
not obtained (if there were any,
they would be discarded as noise).  One finds for the counting rate
$$
{\Delta N(t_i)\over\Delta t}\approx 0,\qquad\qquad t<0\leqno(4.3)
$$
This is so obvious that one usually does not mention it.  For $t>0$ one can
fit the experimental counting
rate with the exponential function to as good accuracy as one wants (by
taking larger $N$ and smaller
$\Delta t$):
$$
{\Delta N(t_i)\over\Delta t}\approx Ne^{-\Gamma t}\qquad t>0=t_0\leqno(4.4)
$$
The $\approx$ in (4.4) means, as in (2.1), the 
equality between experimental
numbers and the idealized, 
theoretical hypothesis $e^{-\Gamma t}$ [29]. 
%%%%%%%%%%%%%%%%%%%%%%%%%%%%%%
%Therefore any infinitesimal (not
%given in terms of the scale
%${1\over\Gamma}$) deviations from the exponential law that are derived from
%mathematical idealizations
%(e.g., the Hilbert space idealization) are physically meaningless.
%%%%%%%%%%%%%%%%%%%%%%%%%%%%%%%%%%%%%%%%%%%%%

Theoretically, the counting rate is given by the probability rate
$$
R(t)={{d{\cal P}(t)}\over{dt}}\leqno(4.5)
$$
where ${\cal P}(t)$ is the probability
for the observable $\Lambda(t)$
of (4.2) (i.e., 
$\pi^+\pi^-$) in the state $W^{K^{\circ}}$.

According to the postulate (2.17), the probability should be given by,
$$
{\cal P}(t)={\rm Tr}(\Lambda(t)W^{K^o})={\rm Tr}(\Lambda W^{K^o}(t))
\ {\rm for}\ t\geq t_0=0
\leqno(4.6)
$$
$$
{\rm where}\qquad\qquad 
W^{K^{\circ}}(t)=e^{-iHt}W^{K^{\circ}e^{iHt}}\quad{\rm for}\quad  t\geq
t_0\leqno(4.7)
$$
For $t<t_0=0$, $W^{K^{\circ}}(t)$ is nonexistent because
the $K^{\circ}$ had not been prepared by $t<t_0$.

To calculate theoretical results
%%%%%%%%%%%%%%%%%%%%%%%%%%%%%%%%%%%%%%%%%%%
%\footnote{$^6$}{Figure (XXX) 
%shows the experimental counting rate without subtraction of a
%theoretical interference term attributed to $K_L$, which leads to the
%deviations
%around $t=12\tau_s$ but not for small values of $t$.  In Fig.~(YYY) 
%(which is from a
%later experiment) this interference term has been subtracted.} 
%%%%%%%%%%%%%%%%%%%%%%%%%%%%%%%%%%%%%%%%%%%%%
 that agree with the observations (4.3) and
(4.4), one has to choose the
state operator $W^{K^\circ}(t)$ in (4.6) such that $W^{K^\circ}(t)$ is
nonexistent for 
$t<t_0=0,$ and such that for
$t>t_0=0$, yields by (4.5) and (4.6), a result that is in
agreement with the right hand side of (4.4).  The
state operator which has this property is given by (2.16),
$$
W^{K^\circ}(t)=|F(t)\rangle\langle F(t)|\leqno(4.8)
$$
where 
$$
F(t)=U_+^\times(t)F_0\leqno(4.9)
$$
is a semigroup solution (2.15) of the
quantum mechanical Cauchy problem, and where the initial vector is
given by the Gamow vector
$$
F_0=|E_{\rm R}-i\Gamma/2^-\rangle\in\Phi_+^\times\leqno(4.10)
$$
with $E_{\rm R}=m_S$ and $\Gamma={{1}\over{\tau_S}}$ for the $K^\circ$
at rest [30].

Then we
obtain the time evolved state vector (3.20) by applying the semigroup (3.15).
For this vector $|E_{\rm R}-i\Gamma/2^-\rangle$ (and only for this
Gamow vector$|E_{\rm R}-i\Gamma/2^-\rangle\in\Phi^\times_+$, which is 
defined by the pole term of the $S$-matrix) one derives the exact Golden
Rule with the (exact)
exponential decay low (3.23), thus reproducing the right hand side of (4.4).

Therewith we see that the Gamow state vector $\psi^{\rm
G}=|E_{R}-i{\Gamma\over2}^-\rangle$ or the operator
$W^{\rm G}=|\psi^{\rm G}\rangle\langle\psi^{\rm G}|$, whose time evolution
is governed by the exact
Hamiltonian $H$, describes the decaying neutral Kaon system (4.1) in its
rest frame if $E_R=m_s$ and
$\Gamma\!=\!\Gamma_s\!=\!~{1\over\tau_s},\ W^{\rm G}=W^{K^{\circ}}$.

For this Gamow state one can calculate the decay rate and 
decay probability as a
function of time and obtain
the exponential law for $t>t_0=0$.  The decay probability is the a priory
probability for a single
decaying microsystem $K^{\circ}$ that has been created in the state
$W^{K^{\circ}}$ at the initial time
$t=0$ (for the quantum
mechanical Cauchy problem with semigroup evolution). 
This is the same point of view mentioned at the
beginning of this section for the quantum state of our universe [3],
except that its initial state $\rho(t_0)$ is probably not a ``pure''
Gamow state. Alternatively
$W^{K^\circ}(t)$ can also be thought of as describing the state of an
ensemble of single microsystems
$K^{\circ}$ created at an ``ensemble'' of times $t_0$, all of which are
chosen to be the initial time $t=0$ for the quantum mechanical Cauchy
problem.  Then the decay probabilities are the statistical
probabilities for this ensemble of individual $K^\circ$ systems, but
$t$ in $W^{K^\circ}(t)$ is the time in the 
``life" of each single decaying
$K^{\circ}$-system which had started at $t=0$.
It is not the time in the experimentalists life or the time in the
laboratory or the time of a
``wave-packet'' of $K^{\circ}$'s.

With this interpretation the single quasi-stable particle and the single
quantum universe are perfectly
analogous, and the time $t_0$, at which the preparation of the state is
completed and at which the registration of the observables can begin,
has been observationally defined.
\vfill\eject
\noindent{\bf 5.  Summary and Conclusion}

\n If we want to have a quantum theory that applies to the
(closed) universe as a whole then we would like this
quantum theory to be time asymmetric, because of
the cosmological arrow of time. By the same reasoning
if a quantum theory is to apply to the electromagnetic
field then it should be time asymmetric, because of 
the radiation arrow of time. Standard quantum theory
is time symmetric. This is a mathematical consequence
due to the property of the Hilbert space postulates.

There is a mathematical theory that describes time
symmetric as well as time asymmetric quantum physics.
It is an extension of the Rigged Hilbert Space (RHS)
formulation of quantum mechanics which about
1965 gave a mathematical justification to Dirac's
kets and his continuous basis vector expansion.

To incorporate causality, the RHS theory distinguishes 
meticulously between states and observables for which 
it uses two RHS's $\Phi_{\mp}\subset {\cal H}\subset
\Phi_{\mp}^{\times}$ of Hardy class with complementary
analyticity properties. The dual spaces in the RHS's 
contain, besides the Dirac 
kets $|E^{\pm}\rangle\in\Phi_{\mp}^{\times}$
(${\rm in}\choose{\rm out}$ plane waves), also
Gamow kets $|E_{R}-i\Gamma/2^{\pm}\rangle\in
\Phi_{\mp}^{\times}$. The Gamow kets have all the 
properties attributed to eigenvectors of complex
finite-dimensional Hamiltonian in the phenomenological
effective theories, in particular an exponential time 
evolution and a Breit-Wigner energy wave function. Neither of this
is possible in Hilbert space.\footnote{$^1$}{The Breit-Wigner energy wave 
function would not be in the domain of the Hamiltonian.} 
These effective finite dimensional theories can therefore not be considered
approximations [15] of standard quantum mechanics, but they go
beyond it. The RHS formulation is the mathematical theory of which these 
finite dimensional models, e.g., the two dimensional
Lee-Oehme-Yang theory for neutral Kaons
and the Weisskopf-Wigner method, are approximations.

Features of the exact theory which are not already
features of these effective
models and phenomenological methods are : the Breit-Wigner wave function (3.22)
of the Gamow ket which extends over $-\infty_{\rm II}<E<+\infty$
(rather than the values $0\leq E <\infty$),
the background integral in (3.25), and the exact Golden Rule
[26] for the decay probability $P(t)$ from which the decay
rate $R(t)$ with exponential time dependence (3.23) is
obtained by differentiation.
Dirac's Golden Rule for the initial decay rate is
the Born approximation at $t=0$ of this
exact rule for the decay rate $R(t)$.
These are features which one may wellcome or
accept. The most surprising, unwanted and mostly rejected
feature of the exact RHS theory is the semigroup time evolution
(3.20) and (3.21) of the Gamow state, which is a manifestation of
a fundamental quantum mechanical arrow of time.

Decaying Gamow states can be experimentally isolated as
quasistationary microphysical systems if their time
of preparation can be accurately identified. 
The observed decay probabilities (4.3), (4.4) of the neutral Kaon system
have the same features as derived from a Gamow state,
including the time ordering, (4.6). This time ordering is
the same as the postulated time ordering in the probabilities of the 
histories  of the universe considered as a quantum system,
(2.19)-(2.25). Under this
hypothesis the fundamental quantum arrow of time -- expressing the
vague notion of causality -- can be considered subsumed
under the cosmological arrow of time.
%The CP violation experiments for the neutral kaon, though set
%up for a different purpose, provide a good example.
%But every scattering experiment is an example of
%time asymmetry in quantum physics, due to the time 
%asymmetric boundary conditions realizable in our laboratories :\hfil\break
%It is easy to prepare two uncorrelated incoming beams
%that scatter into strongly correlated outgoing 
%spherical waves as done in the typical scattering experiment.
%It is experimentally hopeless to prepare a state consisting
%of two strongly correlated spherical waves (with fixed relative
%phase) in such a way that after the scattering two uncorrelated
%plane waves emerge. The latter would be the time reverse\
%of the setup for a physical experiment. It would have to 
%be so complicated that it is totally improbable to
%accomplish.

\vfil\eject
\n {\bf Acknowledgement}

\n The author would like to acknowledge the very
valuable conversations on many points discussed in this 
paper with M.~Gell-Mann, which were actually the origin of 
this article. He would also like to thank J.~Hartle 
for clarifying remarks on their papers. S.~Wickramasekara
and H.~Kaldass provided suggestions and help with the 
preparation of this article. Support from the Welch Foundation is
gratefully acknowledged.

\vfil\eject
\n {\bf Appendix A}
%\vskip .2cm

\n In the Hilbert space formulation of quantum mechanics,
the linear scalar product space $\Phi_{\rm alg}$ is completed with
respect to the  
norm to obtain a 
Hilbert space $\cal H$. The Hamiltonian $H$ 
in the Schroedinger--von Neumann
equations (2.8) is self-adjoint and semi-bounded, and the initial data 
$\phi_0, \psi_0\in{\cal H}$.

\n Then one has the following mathematical theorems:
\item{1.} (Gleason) For every probability ${\cal P}(\Lambda)$, 
there exists a
positive trace class operator $\rho$ such that ${\cal P}(\Lambda)={\rm
Tr}(\Lambda\rho)$ [31]. 
\item{2.} (Stone--von Neumann) 
The solutions of the Schroedinger--von Neumann equations for this
$\rho$  are time
symmetric and given by the group $U{^\dagger}(t)=e^{-iHt}$ 
of unitary operators [32].
\item{3.} (Hegerfeldt) 
For every Hamiltonian $H$ (self-adjoint, semi-bounded),\hb
either\hb
Tr($\Lambda(t)\rho$)=Tr($\Lambda\rho(t)$)=0 \quad\quad for
$-\infty<t<\infty$\hb
or,\hb 
Tr($\Lambda(t)\rho$)=Tr($\Lambda\rho(t))>0$ \quad\quad for all
$t$ (except on a set of Lebesgue measure zero).\hb
%\rightline{(except on a set of Lebesgue measure zero)\qquad\qquad}
Here, $\Lambda$ can be any positive, self-adjoint operator such as
$\Lambda=|\psi\rangle\langle\psi|$ or $\Lambda=C_\alpha$ of (2.20) 
and $\rho$
any trace class operator like $\rho=|\phi\rangle\langle\phi|$ or
$\rho=\rho(t_{\rm big\ bang})$ [33]. 

Theorem 1 says that all probabilities must be 
given by the trace. From theorem 3 it then follows that there cannot
be a state $\rho$ in the Hilbert space ${\cal H}$ that has been created or
prepared a finite time $t-t_0$ ago, and for which therefore
Tr$(\Lambda\rho(t)=0$ for $t<t_0$, which for $t\geq t_0$ decays into
decay products $\Lambda$ with a decay probability that is different
from zero. This means, there exist no
elements $\phi$ in $\cal H$ 
that can represent decaying states. Also absent are the states
$\rho^{\rm eff}(t_i)$ that have been created at times 
$t=t_0=t_{\rm big\ bang},\ 
t=t_1>t_0,\ t=t_2>t_1,$ etc., and whose probabilities
Tr$(P(t_n)\rho^{\rm eff}(t_{n-1}))$ are different from zero. 
%%%%%%%%%%%%%%%%%%%%%%%%%%%%%%%%%%
%This excludes any mathematical theory of decay in the Hilbert space
%formulation of quantum mechanics (for which the decaying state $\rho$ 
%is not already decaying even before it was prepared).\hb
%%%%%%%%%%%%%%%%%%%%%%%%%%%%%%%%%%%%%%%%%%%%%%%%%%%%%%%%
Theorem~2 prohibits the asymmetric time evolution of a state in $\cal
H$ and therewith the existence   
of a distinguished time $t_0$ of creation.                              
\vfill\eject

\noindent{\bf Appendix B: $S$ Matrix and Lippmann-Schwinger Equation}

\n Every experiment consists of a preparation stage and a registration
stage. In the preparation stage of the scattering experiment, a 
(mixture of) initial states $\phi^{\rm in}$ is prepared before the 
interaction $V=H-K$ is effective (e.g., by an accelerator 
outside the interaction region of the target). 
The initial state vectors $\phi^{\rm in}$, describing the non-interacting beam 
and target, evolve in time according to the free Hamiltonian 
$K$: $\phi^{\rm in}(t)=e^{-iKt}\phi^{\rm in}$. When the beam reaches the 
interaction region, the free in-state $\phi^{\rm in}$ turns into the exact 
state vector $\phi^{+}$ whose time evolution is governed by 
the exact Hamiltonian $H=K+V$
$$\Omega^{+}\phi^{\rm in}(t)\equiv \phi^{+}(t)=e^{-iHt}\phi^{+}
=\Omega^{-}\phi^{\rm out}\leqno(B.1)$$
This vector leaves the interaction region and ends up as the well determined 
state $\phi^{\rm out}$. The state vector $\phi^{\rm out}$ is determined from 
$\phi^{\rm in}$ by the dynamics of the scattering process:
$$\phi^{\rm out}=S\phi^{\rm in}\qquad S=\Omega^{-\dagger}
\Omega^{+}\leqno(B.2)$$
$\phi^{\rm in}$ {\it is controlled and determined by the preparation 
apparatus}. $\phi^{\rm out}$ is also controlled by the preparation 
apparatus and is in addition determined by the interaction $V$.

In the registration stage, the detector outside the interaction
region does not detect $\phi^{\rm out}$, but
rather it detects an observable $\psi^{\rm out}(t)=e^{iKt}\psi^{\rm out}$
(or a mixture thereof). $\psi^{\rm out}$~{\it is controlled by the 
registration apparatus} (trigger, energy efficiency, etc., of the 
detector). The detector counts are a measure of the probability
to find the observable (property) $|\psi^{\rm out}\rangle\langle\psi
^{\rm out}|$ in the state $\phi^{\rm out}$. This probability $|(\psi^{\rm out},
\phi^{\rm out})|^{2}$ is calculated by the $S$-matrix.

The $S$-matrix is the probability amplitude $(\psi^{\rm out},\phi^{\rm out})$
which is calculated in the following way: 
$$\eqalign{
(\psi^{\rm out},\phi^{\rm out})&=(\psi^{\rm out},S\phi^{\rm in})
=(\Omega^{-}\psi^{\rm out},\Omega^{+}\phi^{\rm in})\cr
&=(\psi^{-},\phi^{+})=\int_{0}^{\infty}dE\langle\psi^{-}|E^{-}\rangle S(E+i0)
\langle^{+}E|\phi^{+}\rangle}\leqno(B.3)$$
$\phi^{+}(t)=e^{iHt/\hbar}\phi^{+}$ comes from the prepared 
in-state 
\hbox{$\phi^{\rm in}(t\rightarrow -\infty)
=(\Omega^{+})^{-1}\phi^{+}(t\rightarrow -\infty)$.}
The free observable vector $\psi^{\rm out}$ emerges from the observable vector 
$\psi^{-}$ whose time evolution is governed by the exact 
Hamiltonian $H$.\hfil\break
$\psi^{-}(t)=e^{iHt/\hbar}\psi^{-}$ goes into the measured 
out-state \hbox{$\psi^{out}(t\rightarrow +\infty)
=(\Omega^{-})^{-1}\psi^{-}(t\rightarrow +\infty)$}.
$\Omega^{+}$ and $\Omega^{-}$ are the M{\o}ller wave operators. The
Lippmann-Schwinger equation relates the (known) eigenvectors of the
free Hamiltonian $K$ to two sets of eigenvectors of the exact 
Hamiltonian $H$
$$
\eqalign{
|E^{\pm}\rangle
&=|E\rangle+{{1}\over{E-K\pm i\epsilon}}V|E^{\pm}\rangle\cr
&=|E\rangle+
{{1}\over{E-H\pm i\epsilon}}V|E\rangle
=\Omega^{\pm}|E\rangle}\leqno(B.4)$$
where
$$K|E\rangle=E|E\rangle\quad,\quad H|E^{\pm}\rangle=E|E^{\pm}\rangle\leqno(B.5)
$$
This defines the exact energy wavefunctions in terms of the in- and out-
energy wave functions, whose modulus gives the energy resolution of the
experimental apparatuses: 
$$
\eqalign{(B.6)\,\,\langle^{+}E|\phi^{+}\rangle
=\langle E|\phi^{\rm in}\rangle \,\,\,&
{\rm is\ the\ incident\ beam\ resolution,
\ it\ describes\ the\ energy}\cr
&{\rm 
distribution\ given\ by\ the\ accelerator\ (preparation\ apparatus}).}
$$
$$\eqalign{
(B.7)\,\,\langle^{-}E|\phi^{-}\rangle
=\langle E|\phi^{\rm out}\rangle \,\,\,&
{\rm is\ the\ energy\ distribution\ of\ the\
detected\ state,\ it\ is\ given\ by}\cr
&{\rm the\ energy\ resolution\ of\ the\ detector\
(registration\ apparatus).}}$$

\n Since $\phi^{\rm in}$ is controlled by the preparation apparatus,
so is $\phi^{+}$. Likewise, since $\psi^{\rm out}$ is controlled
by the registration apparatus, so is $\psi^{-}$.
All this is quite standard, cf.~[34] chapter~$7$,
except that of the two versions, mentioned on p.$188$ of [34]
as equally valid descriptions, we allow only the first version
which is in agreement with our physical intuition of
causality. In order to do this we distinguish between
the set of in-state vectors
\hbox{$\{\phi^{+}\}\equiv\Phi_{-}$} and the set of out-observable vectors
$\{\psi^{-}\}\equiv \Phi_{+}$. This hypothesis is 
quite natural since the state $\phi^{+}$ (or~$\phi^{\rm in}$) must be prepared
before the observable $|\psi^{-}\rangle\langle\psi^{-}|$ 
(or $\psi^{\rm out}$) can be measured in it. As shall be discussed in
section~3, $\Phi_{-}$ and $\Phi_{+}$ are different dense subspaces of 
the same Hilbert space $\cal H$.

\vfill\eject
\centerline{\bf Footnotes and References}

\item{[1]}
The literature on resonances and decay is so large that it is difficult 
to list here 
even a representative selection.  The standard monograph is
M.~L.~Goldberger, K.~M.~Watson, Collision Theory, Wiley, New York (1964).
The irreversible character of quantum mechanical
decay has rarely even been mentioned (exceptions are
C.~Cohen-Tannoudji, et al., {\it Quantum Mechanics}, Vol.~II, p.~1345;
Wiley, New York (1977); T.~D.~Lee, [16] below)
%\vskip7pt
and to our knowledge not been incorporated in a theory of decay.

\item{[2]}
We are {\it not} concerned here with irreversibility in the
quantum theory of open systems for which the asymmetric time
evolution is described by a Liouville equation containing terms
for the effects of the external reservoir. Cf.~footnote 1 of section~2. 

\item{[3]}
M.~Gell-Mann, J.~B.~Hartle, p.~311 in {\it Physical Origins of
Time Asymmetry}, J.~J.~Halliwell, et al., eds., (Cambridge
University, 1994).

\item{[4]}
B.~A.~Lippmann, J.~Schwinger, Phys.~Rev., {\bf 79}, 469 (1950);\hb
 M.~Gell-Mann, H.~L.~Goldberger, Phys.~Rev., {\bf 91}, 398 (1953).

\item{[5]}
R.~Ritz, Physikalische Zeitschrift, {\bf 9} (1908) 903; R.~Ritz,
Physikalische Zeitschrift, {\bf 10} (1909) 224; R.~Ritz and
A.~Einstein, Physikalische Zeitschrift, {\bf 10} (1909) 323.

\item{[6]}
J.~von Neumann, {\it Mathematische Grundlagen der Quantentheorie},
(Springer, Berlin, 1931) (English translation. 
Princeton University Press, Princeton, 1955).

\item{[7]}
P.~A.~M.~Dirac, {\it The Principles of Quantum Mechanics},
(Clarendon Press, Oxford, 1930).

\item{[8]}
E.~Roberts, J.~Math.~Phys., {\bf 7} (1966) 1097; A.~Bohm,
{\it Boulder Lectures in Theoretical Physics 1966}, Vol. 9A, (Gordon
and Breach, New York, 1967); J.~P.~Antoine, J.~Math.~Phys., {\bf
10} (1969) 53; 10 (1969) 2276; See also O.~Melsheimer, J.~Math.~Phys.,
{\bf 15} (1974) 902; 917. 

\item{[9]}
A.~Bohm, J.~Math.~Phys., {\bf 22} (1981) 2813; Lett.~Math.~Phys.
{\bf 3} 455 (1978).

\item{[10]}
Some (e.g.\ A.~Pais, CP Violation: the First 25 Years, in {\it CP
Violation}, J.~Tran, Thanh~Van, Eds., Editions Frontiers, (1990))
reserve the name ``particle'' for an object with a unique lifetime (in
addition to the unique mass), in distinction to such superpositions as
$|K^\circ\rangle$ and $|{\bar{K}}^\circ\rangle$ which are the states
in which the neutral Kaon system is prepared. In our theory, these
exact prepared states like the $|K^\circ\rangle$ are the $\phi^+$ of
(3.25), which, in addition to the quasi-stable particle states like
$\psi^{\rm G}_i$ representing 
$K_i^{\rm o}$, also contain a background integral.

\item{[11]}
V.~Weisskopf and E.~P.~Wigner,  Z.~f.~Physik, {\bf 63}, 54 (1930);
{\bf 65}, 18 (1930); W.~Heitler, {\it  Quantum Theory of Radiation},
Oxford (1954).

\item{[12]}
T.~D.~Lee, R.~Oehme and C.~N.~Yang, Phys.~Rev., {\bf 106} (1957)
340.

\item{[13]}
{\it Non-diagonalizable\/} complex Hamiltonian matrices (Jordan blocks) 
led to Jordan vectors instead of eigenvectors.
Jordan vectors have a non-exponential time evolution of the
order of $\hbar / \Gamma$ 
(and are therefore ruled out as decaying vectors). H.~Banmg\"artel
{\it Analytic Perturbation Theory for Matrices and Operators.}
Operator theory. Vol. 15. Birkh\"auser Basel (1985), chapter 2;
P.~V.~Ruuskanen,
Nucl.~Phys.~B{\bf 22}, 253 (1970); E.~Katznelson, J.~Math.~Phys.~{\bf
21}, 1393 (1980); E.~Hernandez and A.~Mondragon, Phys.~Lett.~B{\bf
326}, 1 (1994); A.~Mondragon and E.~Hernandez, J.~Phys.~A{\bf 26},
5595 (1993). 
However, the Jordan block Hamiltonians can also be shown to 
result from a truncation
of the exact Rigged Hilbert Space theory to an effective theory. Its
higher order Gamow states, which correspond to higher order $S$-matrix
poles, and are not representable by vectors but by state operators
have an exponential asymmetric time evolution.
A.~Bohm, M.~Loewe, S.~Maxson, P.~Patuleanu, C.~P\"untmann and
M.~Gadella, J.~Math.~Phys., {\bf 38}, 6072 (1997).

\item{[14]}
E.~Maglione, L.~S.~Ferreira and R.~J.~Liotta, 
Phys.~Rev.~Lett., {\bf 81} (1998) 538;
O.~I.~Tolstikhin, V.~N.~Ostrovsky and H.~Nakamura, Phys.~Rev.\ 
A{\bf 58}, 2077 (1998);
H.~Hogreve, Phys.~Lett.\ A{\bf 201}, 111 (1995);
L.~S.~Ferreira in {\it Resonances}, E.~Br\"andas et~al.~[Eds]
Lecture Notes in Physics Vol.~{\bf 325} p.201, Springer Berlin (1989).

\item{[15]}
M.~Levy, Nuovo Cimento,  {\bf 13}, 115 (1959).

%\item{[15]}N.~G.~van Kampen
\item{[16]}
T.~D.~Lee, Chapter 13 of {\it Particle Physics and Introduction to
Field Theory}, Harwood Academic, New York, 1981.  In
this reference the quantum mechanical time reversed state is
called complicated and improbable.

\item{[17]}
A.~Kossakowski, On Dynamical Semigroups and Open Systems, in
{\it Irreversibility and Causality},
A.~Bohm, H.~D.~Doebner, P.~Kielanowski [Eds.] Springer, Berlin
(1998), p.~59, where further references to this irreversibility
of quantum statistics can be found.

\item{[18]}
G.~Ludwig, {\it Foundations of Quantum Mechanics}, Volume I,
Springer-Verlag, Berlin, (1983) and Volume II, (1985); {\it An
Axiomatic Basis of Quantum Mechanics}, Volume I,
Springer-Verlag, Berlin, (1983) and Volume II, (1987).

\item{[19]}
M.~Gell-Mann and J.~B.~Hartle, in {\it Complexity, Entropy and the
Physics of Information}, SFI Studies in Science and Complexity
Vol.~VIII, W.~Zurek, Ed., (1990).

\item{[20]}
M.~Gell-Mann, J.~B.~Hartle, {\it Proceedings, 4th Drexel  Symposium
on Quantum Non-Integrability,} preprint UCSBTH-95-28.

\item{[21]}
A.~Bohm and M.~Gadella, {\it Dirac Kets, Gamow Vectors and Gel'fand
Triplets}, Lecture Notes in Physics, {\bf 348},
Springer-Verlag, Berlin, (1989).

\item{[22]}
The Gamow vectors were envisioned by G.~Gamow (Z.~Physik, {\bf 51},
204 (1928)) as far back as the inception of the Dirac
kets. Mathematically, they were no more ill defined than the Dirac
kets, though they never acquired the popularity of the latter.  The
reason was that, although for Dirac kets 
at least the probability density is
finite everywhere, the probability density of Gamow's
wavefunctions $|\langle r,\theta,\varphi|\psi^{\rm G}(t)\rangle|^2$ 
increased exponentially
for large values of the distance $r$ and large negative values of time
$t$, (the ``exponential catastrophe''). The origin of this exponential
catastrophe is, that the emission of decay products
had been assumed to go on for an arbitrarily long time, $t\to-\infty$, as
dictated by the unitary group $e^{-iHt},\ -\infty<t<\infty$. In
reality, the emission of decay products must have begun at some finite time
$t=t_0=0$ in the past, and the time evolution of the decaying state,
as we now know, is described by the semigroup $e^{-iH^{\times}t},\
0<t<\infty$. Therefore, in the time asymmetric
theory the probability density $|\langle
r,\theta,\varphi|\psi^{\rm G}(t)\rangle|^2$ is only defined for times
$t>t_0(r)={{r}\over{v}}={{mr}\over{p}}\approx {{mr}\over{\sqrt{2mE_R}}},$
 and for these times one can show that there exists 
no exponential catastrophe 
for the probability densities. Thus the semigroup evolution of the
Gamow vectors is not only a consequence of the mathematical theory,
but also a necessity for the physical interpretation of the position
probabilities $\int_{\Delta{\vec r}}r^2drd\theta d\varphi|\langle
r,\theta,\varphi|\psi^{\rm G}(t)\rangle|^2$.

\item{[23]}
The function $G_{+}(E)\left(=\langle^{-}E|\psi^{-}\rangle
=\langle E|\psi^{\rm out}\rangle\right)$ is a very well
behaved function of the upper half
plane ${I\kern-.55em C}^{+}$ if it is well behaved, 
i.e.~$G_{+}(E)\in {\cal S}$ (Schwartz space) and if 
it is the boundary value of an analytic
function $G_{+}(z)$ in the upper half plane ${I\kern-.55em C}^{+}$
which vanishes faster than any power at the infinite
semicircle (i.e.~$G_{+}(E)\in {\cal H}_{+}$). Similarly the function 
$G_{-}(E)\left(=\langle^{+}E|\phi^{+}\rangle=\langle E|\phi^{\rm in}
\rangle\right)$ is a very well behaved function of the lower half
plane ${I\kern-.55em C}^{-}$ if it is well behaved ($G_{-}(E)\in{\cal S}$)
and if it is the boundary value of an analytic function in 
the lower half plane ${I\kern-.55em C}^{-}$ which vanishes 
sufficiently fast
at the lower semicircle ($G_{-}(E)\in {\cal H}_{-}$). 
For the definition of Hardy
class functions and their mathematical properties needed here
see Appendix~A.2 of reference [25] and
P.~L.~Duren, {\it ${\cal H}^p$ Spaces}, Academic Press, New York,
(1970).

\item{[24]}
A.~Bohm, I.~Antoniou, P.~Kielanowski, Phys.~Lett.~A{\bf 189}
442 (1994); A.~Bohm, I.~Antoniou, P.~Kielanowski, J.~Math.~Phys., {\bf
36}, 2593 (1995).

\item{[25]}
A.~Bohm, S.~Maxson, M.~Loewe, M.~Gadella, Physica A{\bf 236},
485 (1997).

\item{[26]}
A.~Bohm, N.~L.~Harshmann in {\it Irreversibility and Causality,}
p.~225, 
Sect.~7.4; A.~Bohm, H.~D.~Doebner, P.~Kielanowski [Eds.]
Springer, Berlin (1998).

\item{[27]}
For a detailed analysis of the $K_L-K_S$ system in the framework
of time asymmetric quantum mechanics with and without CP
violation, see A.~Bohm; hep-th/970542.

\item{[28]}
This is the simplified schematic diagram of several generations
of experiments measuring CP violation in the neutral Kaon
system; J.~Christenson, J.~Cronin, V.~Fitch, R.~Turley,
Phys.~Rev.~Lett., {\bf 13}, 138 (1964);
K.~Kleinknecht, {\it CP Violation}, p.~41, C.~Jarskog (Ed.),
World Scientific (1989) and references therein; NA{\bf 31},
G.~D.~Barr et al., Phys.~Lett.~B{\bf 317}, 233 (1993);
E731; L.~K.~Gibbons, et al., Phys.~Rev.~Lett., {\bf 70},
1199 and 1203 (1993).
%%%%%%%%%%%%%%%%%%%%%%%%%%%%%%%%%%%%%%%%%%%%%%%%%%%%%%%%%%%%%%%%
%\item{[28]}
%The $K^\circ$ is a relativistic decaying system and our
%discussion here is in terms of the non-relativistic Gamow
%vectors.  Relativistic Gamow kets can be defined from the poles
%of the relativistic $S$-matrix at the value of the invariant
%mass square $\Delta=\Delta_R=(m_R-i{\Gamma\over2})^2$ which
%have at rest the same semigroup time evolution (3.20) with
%$E_R\to m$ and $\Gamma\to{\hbar\over\tau_R}$.  A. Bohm, H.
%Kaldass, et al. in preparation.
%%%%%%%%%%%%%%%%%%%%%%%%%%%%%%%%%%%%%%%%%%%%%%%%%%%%%%%%%%%%%%%%%
\item{[29]}
This approximate agreement between a sequence of rational
numbers $\Delta N(t)\over \Delta t$ on the experimental 
side and a continuous function of real numbers $e^{-\Gamma t}$
on the theoretical side is the fundamental limit to which
the exponential law can be verified; experimental limitations
given by the resolution, e.g., the finite size of $\Delta t$, are still
more important. Therefore any {\it infinitesimal} (not given
in terms of the scale ${1}\over{\Gamma}$) deviations from the 
exponential law that are derived from a mathematical theory
(e.g., the Hilbert space idealization, L.~A.~Khalfin, JETP Lett.,{\bf 15},
388 (1972)), are physically meaningless. 
More important for observed deviations from 
the 
exponential law, 
(e.g.\ S.~R.~Wilkinson et.\ al.,\ Nature {\bf 387}, 575 (1997)),
are the limitations on the preparation side of the 
experiment. According to (3.25) the prepared state $\phi^{+}$
contains, in addition to the exponentially decaying Gamow
state $\psi^{G}$ (assuming $N=1$), the background integral,
which does not have exponential time evolution. Theoretically,
the background term 
could be infinitesimally small, but since it 
depends upon the state preparation
i.e., the function $\langle ^{+}E|\phi^{+}\rangle=\phi^{\rm in}(E)$,
$0\leq E<\infty$, the effect of the background term can be
substantial. Experimental conditions may thus not allow the 
isolation of the Gamow state $\psi^{\rm G}$ 
from this background
term, and consequently deviations from the exponential law
will be observed. This is a familiar effect in resonance scattering 
experiments where deviations from the pure Breit-Wigner distribution
of the Gamow state are observed and attributed to the
background phase shifts, see e.g., A.~Bohm, {\it Quantum
Mechanics,} 3$^{\rm rd}$ ed., sections XVIII.6--9, and XX.3,
Springer-Verlag, New York, (1994). It is not experimental evidence 
against the existence of a Gamow state with Breit-Wigner 
energy distribution
and exponential time evolution.

\item{[30]}
The $K^\circ$ is a relativistic decaying system and our
discussion here is in terms of the non-relativistic Gamow
vectors.  Relativistic Gamow kets can be defined from the poles
of the relativistic $S$-matrix at the value of the invariant
mass square $s_{\rm R}=(m_R-i{\Gamma\over2})^2$. They
have at rest the same semigroup time evolution as (3.20) with
$E_R\to m$ and $\Gamma\to{\hbar\over\tau_R}$.  A.~Bohm, H.~Kaldass, et
al., (in preparation).

%\item{[A.1]}
\item{[31]}
A.~M.~Gleason, J.~Math., {\bf 6}, 885 (1957)

%\item{[A.2]}
\item{[32]}
M.~H.~Stone, Ann.~of Math. {\bf 33}, 643 (1932)

%\item{[A.3]}
\item{[33]}
G.~C.~Hegerfeldt, Phys.~Rev.~Lett. {\bf 72}, 596 (1994)

%\item{[B.1]}
\item{[34]}
R.~G.~Newton, {\it Scattering Theory of Waves and Particles} 
(McGraw-Hill, 1966) 

\vfill\eject

%%%%%%%%%%%%%%%%%%%%%%%%%%  Figures
\vfill\eject
\input epsf.tex
\vskip 3 true cm
\epsfysize = 4 true in
\centerline{\epsffile {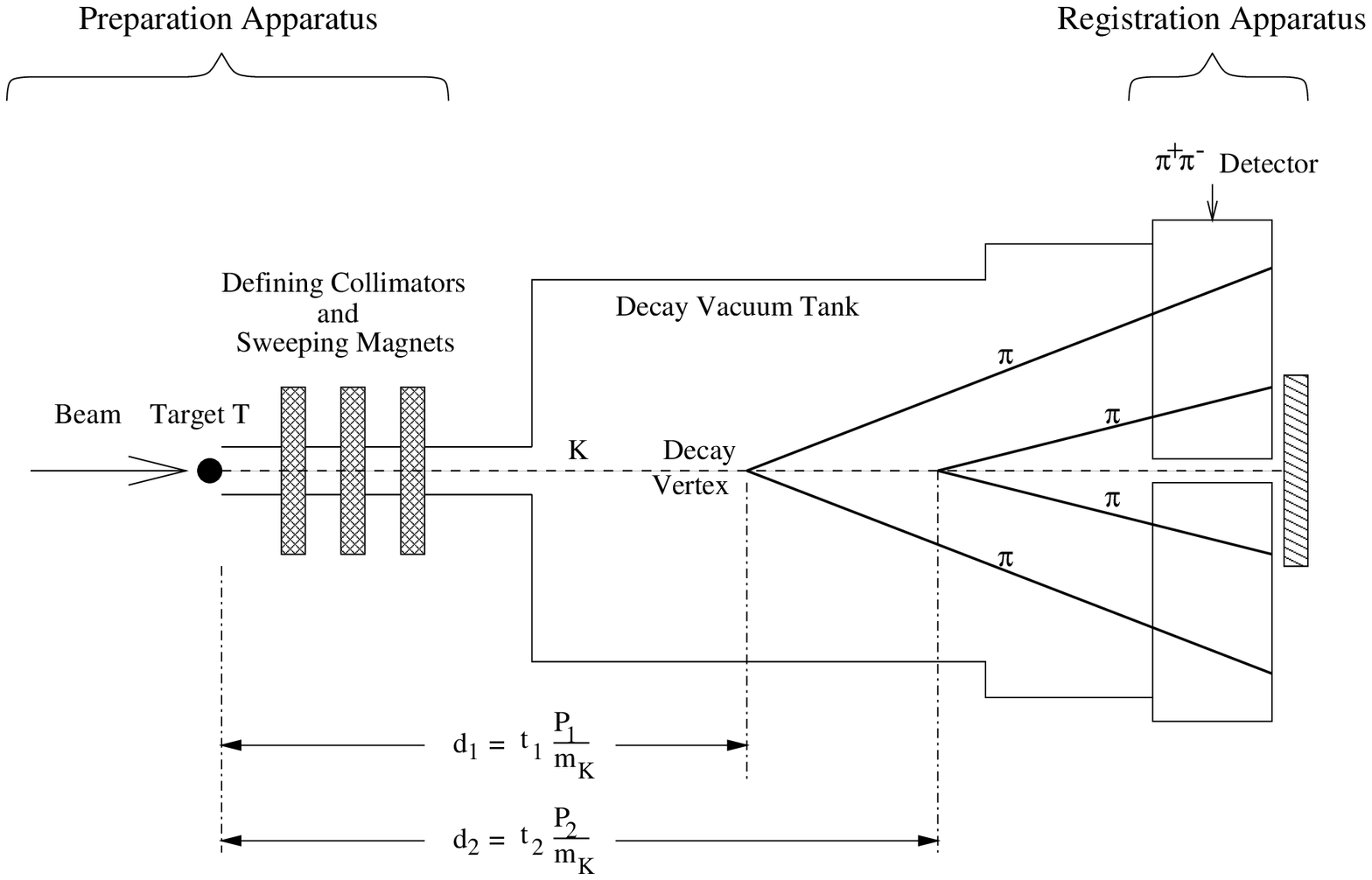}}
\vskip 0.3 true cm
\line{\hfil Figure 1 Schematic diagram of the neutral K-meson decay 
experiment \hfil}
\vskip 1 true cm
\epsfysize = 3.3 true in
\centerline{\epsffile {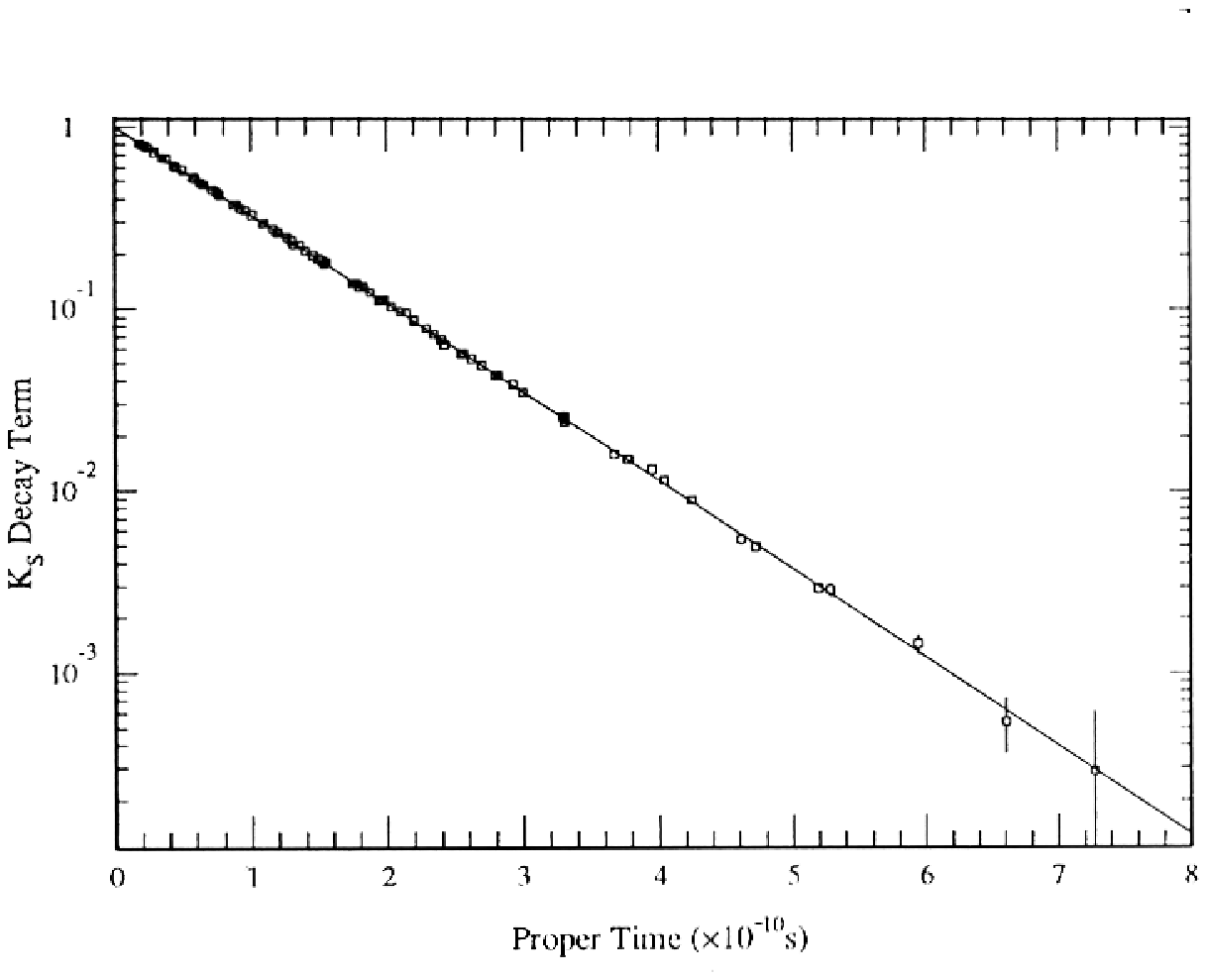}}
%\vskip 0.3 true cm
\line{\hfil Figure 2 ${\rm K}_{\rm S}$ decay versus proper time \hfil}

%%%%%%%%%%%%%%%%%%%%%%%%%%%%%%%%%%%%%%%%%%%%%%%%%%%%%

\vfil\eject\bye